# Capillary-driven binding of thin triangular prisms at fluid interfaces

Joseph A. Ferrar[1], Deshpreet S. Bedi[2,3], Shangnan Zhou[2,4], Peijun Zhu[2,5], Xiaoming Mao[2]*, and Michael J. Solomon[1]*

[1] Department of Chemical Engineering, University of Michigan, Ann Arbor, MI 48109

[2] Department of Physics, University of Michigan, Ann Arbor, MI 48109

[3] Martin Fisher School of Physics, Brandeis University, Waltham, MA 02454

[4] Theoretical Physics and Department of Physics, Stanford University, Stanford, CA 94305

[5] Department of Physics and Astronomy, University of Pittsburgh, Pittsburgh, Pennsylvania 15260

(*Corresponding Author: maox@umich.edu and mjsolo@umich.edu)

**Abstract**

We observe capillary-driven binding between thin, equilateral triangular prisms at a flat air-water interface. The edge length of the equilateral triangle face is 120 μm, and the thickness of the prism is varied between 2 and 20 μm. For thickness to length (T/L) ratios of 1/10 or less, pairs of triangles preferentially bind in either a tip-to-tip or tip-to-midpoint edge configurations; for pairs of particles of thickness T/L = 1/5, the tip of one triangle binds to any position along the other triangle's edge. The distinct binding configurations for small T/L ratios result from physical bowing of the prisms, a property that arises during their fabrication. When bowed prisms are placed at the air-water interface, two distinct polarity states arise: prisms either sit with their center of mass above or below the interface. The interface pins to the edge of the prism's concave face, resulting in an interface profile that is similar to that of a capillary hexapole, but with important deviations close to the particle that enable directed binding. We present corresponding theoretical and numerical analysis of the capillary interactions between these prisms and show how particle bowing and contact-line pinning yield a capillary hexapole-like interaction that results in the two sets of distinct, highly-directional binding events. Prisms of all T/L ratios self-assemble into space-spanning open networks; the results suggest design parameters for the fabrication of building blocks of ordered open structures such as the Kagome lattice.

**Introduction**

Attractive, long-range capillary interactions arise between particles at an air-liquid or liquid-liquid interface because they minimize the free energy generated by the particle-induced curvature of the interface.[1–4] Self-assembly of colloidal, granular, and millimeter-scale particles has been observed at both air-water and oil-water interfaces due to such capillary-induced pair attractions. At the colloidal scale, short-range electrostatic repulsions also influence self-assembly.[5–7] Recently, spatially anisotropic capillary attractions have been used to produce ordered particle chains and complex open networks at fluid interfaces. For example, colloidal ellipsoids at oil-water and air-water interfaces form such structures. Particle configurations that arise at the interface are dependent on particle surface geometry, chemistry, and wettability.[7,8] For example, cylinders and related anisotropic shapes assemble into chains at an oil-water interface, with the specific particle faces that bind determined by the curvature of the particle face. In these cases, the spatial anisotropy of the capillary interaction is a consequence of differences in the local curvature of the particle.[9,10] Specifically, cylindrical particles at an oil-water interface generate an elliptical quadrupolar interaction in the far field: the interface deforms in one direction at the flat ends of the cylinder and the opposite direction at the curved edges. These deformations yield attractive capillary interactions between faces with like-curvature and repulsions between faces with opposite-curvature.[9,10]

Capillary-driven self-assembly therefore is a path to the bottom-up assembly of open and network structures. Such structures are targets for self-assembly due to interesting and potentially useful properties, ranging from photonic bandgaps to unusual mechanical response, that arise from the incorporation of voids into material structures.[11–14] These networks and voids deform in ways that significantly differ from close-packed structures, and can lead to mechanical properties such as negative Poisson's ratio and rigidity at ultra-low density. For example, open networks of colloidal ellipsoids assembled at a fluid-fluid interface exhibited a significantly enhanced low frequency modulus as compared to close-packed networks of colloidal spheres at similar particle concentrations.[7]

Current methods to fabricate open networks include the above described capillary-driven assembly of colloidal ellipsoids,[7,8] and polymer-molded microhexagram prisms,[15] millimeter-scale branched shapes produced by 3D printing,[16] self-assembly of patchy colloidal spheres,[17–19] and top-down approaches on the granular and millimeter-scale such as polymeric 3D-printing,[20] quasi-2D-polymer molding[21] and lithography.[22] Bottom-up self-assembly methods can be advantageous compared to these top-down methods, because of the potential scalability of self-assembly processes.[23,24]

Here we investigate the possibility of using a hexapolar-like interaction generated between pairs of thin, triangular microprisms to self-assemble space-spanning open networks at low particle concentrations. Assembly of such a rigid, stabilizing network by control of lateral interactions could yield complex fluids with useful bulk and interfacial rheological properties of interest in a variety of fields and industries, such as food science, drug delivery, and petrochemical processing.[25–28]

Thin prisms – quasi-2D shapes with finite but small thickness – can generate capillary interactions at fluid-fluid interfaces if sufficient interface deformation is induced at the prism edges. The symmetry of thin, triangular prisms indicates that the interaction will be similar to that of capillary hexapoles when these prisms are not too close. This interaction may lead to binding of the triangles at vertices and yield ordered structures such as the kagome and the twisted kagome lattices – a family of isostatic structures with a unit cell of two inverted triangles (Figure 1). These kagome lattices are known to display unusual mechanical properties such as a negative Poisson's ratio and floppy edge modes.[12,14,17,29–33] To improve the prospects for assembling such complex open structures – either ordered or disordered – the pair-binding behavior of thin homogenous microprisms at interfaces should be investigated. Better understanding of the transient and steady-state binding can identify conditions for which ordered and/or disordered networks (Figure 1) might occur; each structural family might itself exhibit interesting mechanical properties.[7] Open, planar networks – both disordered and ordered – are therefore interesting targets for interfacial self-assembly.

Here we observe capillary-driven binding of thin, triangular prisms, with edge lengths ~120 μm and thicknesses between 2.5 and 20 μm at an air-water interface. The pairwise interaction between prisms is measured and modeled. The particles are produced by polymeric photolithography; the anisotropic, directional interactions are introduced by the unexpected generation of a capillary hexapole, which arises due to the 2D triangular shape and the contact line curvature induced by edgewise bowing of the prisms that is introduced at the time of synthesis. We record the different types of binding events observed between the vertices and flat edges of the interacting prisms. The type of binding event is predictable from the up/down polarity of particle attachment to the interface, which is well characterized by imaging out-of-plane and by environmental scanning electron microscopy. From the particle shape and bowed radius of curvature, we compute the interface geometry and the resulting capillary interaction numerically. We find that the capillary interaction is similar to hexapolar interaction in the far field, but deviates from ideal hexapoles in the near field such that the variability of the potential is largest at the tips. We also simulate trajectories of particle binding events numerically using the potential we calculated, and we obtain good agreement with qualitative features of the experimental results. These results can inform the structural design of complex open networks from interfacial building blocks.

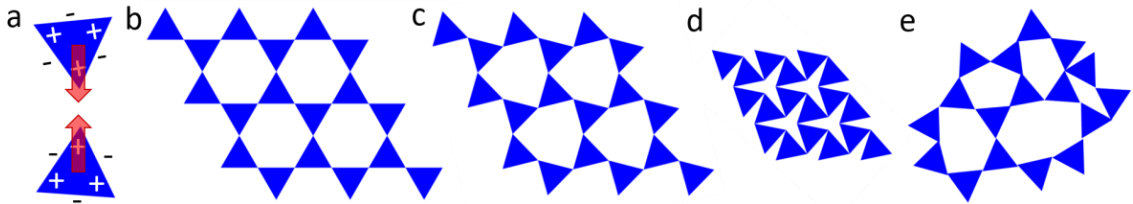

*Figure 1 Hexapole-like capillary interaction between triangles may lead to the self-assembly of kagome lattices. (a) Hexapole-like interactions between triangles (positive at tips and negative at edges) cause tip-to-tip binding. (b) The kagome lattice where edges of triangles form straight lines. (c,d) two twisted kagome lattices with different twisting angle. These different versions of the kagome lattices are related by a soft deformation which only changes the bond angle, which leads to the negative Poisson's ratio of this structure. (e) Depending on the strength of the hexapole-like interaction, disordered assemblies of triangles may also appear.*

**Materials & Methods**

*Particle Fabrication*

Particles are fabricated via SU-8 photolithography.[9,10,34] First, a sacrificial release layer of Omnicoat (Microchem Corp.) is spun onto a glass wafer (D-263 borosilicate glass, Precision Glass & Optics) and baked at 200 C until cured to a thickness of tens of nanometers (1-2 minutes). After cooling to room temperature, SU-8 2000 series photoresist (Microchem Corp.) is spun on top of the Omnicoat layer to the desired prism thickness and baked at 95 C until cured (~2-5 minutes depending on resist thickness). Next, the wafer is exposed to UV light (365 and 405 nm) through a chrome photomask that encodes the particle pattern (Fineline Imaging) until exposure energies of 60-150 mJ (depending on resist thickness) are achieved. The wafer is then heated at 95 C for 2-5 minutes (depending on resist thickness) to ensure adequate cross-linking of the exposed photoresist.

The wafer is immersed in SU-8 developer solution (Microchem Corp.) until the non-photopolymerized SU-8 is washed away (~1-5 minutes depending on resist thickness), leaving the cross-linked particles immobilized on top of the release layer. The wafer is exposed to oxygen plasma for 20 minutes, which facilitates release of the particles into isopropanol. The particles are stored in isopropanol, where they remain stable for several weeks. This process yields approximately $10^6$ particles per fabrication. Figure shows the 4 types of equilateral triangular prisms fabricated. All prisms have an edge length of 120 μm, and thickness of: (a) 2.5 μm, (b) 5 μm, (c) 12 μm, and (d) 20 μm. The ratio of the thickness (T) to length (L) of the prisms is a characteristic parameter; we hereafter refer to each type of prism as: (a) T/L = 1/50, (b) T/L = 1/25, (c) T/L = 1/10, and (d) T/L = 1/5.

*Placement of particles at the air-water interface*

A flat interface is formed between air and deionized water in a chamber (Thermo Scientific Lab-Tek II, 2 Chamber, coverslip 0.13-0.17 μm thick, type 1.5) of dimension 2.0 x 2.0 cm, mounted on to the stage of a Nikon A1Rsi confocal microscope. The chamber's large experimental area and acrylic walls allow for a flat air-water interface to form – without the need for surface modification of the chamber – through careful placement of water in the chamber with a transfer pipette. The walls of the chamber are manually wet prior to filling the center of the chamber with water, in order to prevent uneven attachment of the interface to the walls of the chamber. 10 μL of the particle stock solution is carefully placed in one or two drops at the air-water interface using a gas-tight Hamilton 100 μL syringe.

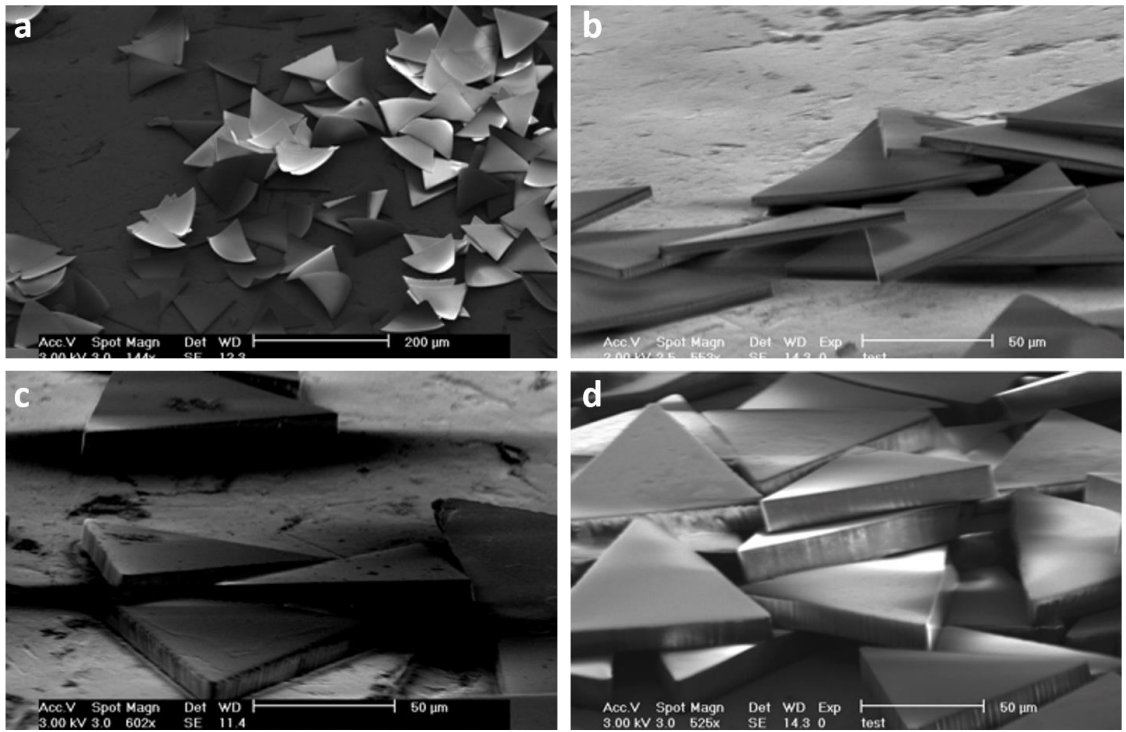

*Figure 2* SEM images of thin, equilateral triangular microprisms from SU-8 epoxy resin. Equilateral triangle (edge length, L =120 μm) prisms of varying thickness (T) a) T ~ 2.5 μm, T/L = 1/50, b) T ~ 5 μm, T/L = 1/25, c) T ~ 12 μm, T/L = 1/10, d) T ~ 20 μm, T/L = 1/5.

*Observation of binding events with optical and reflection microscopy*

The interface is imaged with the transmission and 488 nm reflection channels of a Nikon A1Rsi confocal microscope (10x objective, NA = 0.25) in a square region of 1270 x 1270 μm. Images of pair binding and assembly are acquired at frame rates of 15 frames per second (fps) for prisms of T/L $\geq$ 1/25 and 30 fps for T/L = 1/50. For pair binding experiments, particle positions, relative orientations, and trajectories are tracked by least square fitting of particle edges, as detected by scikit-image (http://scikit-image.org/).

*Quantifying capillary attraction energies through observation of interface deformation with environmental SEM*

Environmental SEM (eSEM, FEI Quanta 3D) is used to observe interface deformation and curvature around the edges of the particles. A gel trapping technique is used to immobilize particles at the interface.[10,35] Briefly, gellan gum, which was generously supplied as a gift from CP Kelco, (low acyl Kelcogel, 2 wt. %) is dissolved in deionized water at 95 °C. The gellan solution remains fluid at temperatures greater than 50 °C. The gellan solution is placed into an eSEM imaging chamber at 70 °C, and prisms are spread at the interface. The imaging chamber is at room temperature, a condition at which the gellan solution crosslinks, immobilizing the prisms for later imaging. Identical prism-prism capillary-driven binding is observed at the gellan solution-air interface as is observed at the pure water-air interface, suggesting that the gellan solution has a negligible effect on the capillary-binding mechanism, consistent with reports of right cylinders at gellan interfaces.[10]

*Modeling of the pairwise interaction potential using Surface Evolver*

In this paper, we consider the capillary interaction potential between triangular prisms, for which an analytic solution to Laplace's equation – especially close to the prisms, where simplifying assumptions cannot be made – is not available (more discussion of capillary interactions can be found in the Supplementary Information). Therefore, we use Surface Evolver, a program widely utilized to model the shape of liquid surfaces and interfaces, to numerically calculate the shape of the interface.[36] The solution is achieved by an algorithmic succession of steps involving gradient and conjugate gradient descent iterations and interface mesh refinements to minimize the interfacial energy subject to specific boundary conditions.

As we discuss in the Results section, we compute the interface shape given a pinned contact line around a bowed equilateral triangle of side length 120 µm. In particular, this triangle is formed by the intersection of three planes containing great circles with a thin spherical shell. Specifying the behavior of the contact lines yields one set of boundary conditions; the far-field boundary condition is that the interface is flat. To allow for the condition of mechanical equilibrium to be satisfied, we do not explicitly fix the height of the far-field boundary, which, in effect, allows for changes in the relative height between the prisms and the equilibrium, unperturbed height of the interface.

In order to generate a potential energy landscape of a pair of interacting triangles, we run Surface Evolver simulations on a regularly-spaced grid in $(r, \theta_1, \theta_2)$ configuration space, where $r$ is the distance between the centers of the two triangles, $\theta_1, \theta_2$ are the orientations of the two triangles (see Figure   for their definitions). The parameter ranges are $132 \ \mu m \leq r \leq 360 \ \mu m$ and $0° \leq \theta_1, \theta_2 < 360°$, with grid spacings of $12 \ \mu m$ in distance and $5°$ in orientation. The actual number of simulations needing to be run is substantially reduced by symmetries inherent in the system. Simulations are run for both particles with the same bowing polarity and opposite bowing polarities; the definitions of bowing and polarity are introduced in the results.

*Computing particle trajectories leading to pair binding*

For a particle moving through a fluid at relatively slow speeds and at a low Reynolds number, Re, the drag force is given by

$$\mathbf{F}_d = -\eta_r \dot{\mathbf{r}}. \tag{1}$$

Analogously, a particle rotating in a fluid at slow speeds experiences a drag torque,

$$\tau_d = -\eta_\theta \dot{\theta}. \tag{2}$$

In these equations, $\eta_r$ and $\eta_\theta$ are the viscous damping coefficients for the center-of-mass and rotational degrees of freedom of the triangular prisms, respectively.

Assuming a quasistatic force balance on the particles, we can equate the corresponding drag and capillary forces to obtain the following system of differential equations of motion for the pair of prisms. This is a valid assumption to make, as both the Reynolds number $Re = \rho v a/\mu$, which is a ratio of inertial forces to viscous forces within a fluid, and the capillary number $Ca = \mu v/\gamma$, which is a ratio of viscous forces to surface tension of an interface, where $\rho$ is the density of the liquid, $v$ is the velocity of the particle, and $\mu$ is the dynamic viscosity of the liquid, are quite small (for a set of characteristic values $\rho = 10^3$ kg/m$^3$, $a = 120$ μm, $\mu = 1.002 \times 10^{-3}$ Pa·s, $\gamma = 72 \times 10^{-3}$ N/m, and $v \sim 4 \times 10^{-4}$ m/s, which is representative of the upper range of velocities observed in the dilute binding events, $Re \approx 0.048$ and $Ca \approx 5.6 \times 10^{-6}$, both of which are small compared to unity), so that both inertia and viscous deformation of the interface can be neglected, as in Refs. 8 and 10. In this case, hydrodynamic interactions can safely be ignored, and the force balance equations are

$$\eta_r \partial_t r(t) = -\partial_r U(\theta_1, \theta_2, r) \tag{3.1}$$

$$\eta_\theta \partial_t \theta_1(t) = -\partial_{\theta_1} U(\theta_1, \theta_2, r) \tag{3.2}$$

$$\eta_\theta \partial_t \theta_2(t) = -\partial_{\theta_2} U(\theta_1, \theta_2, r) \tag{3.3}$$

Discretizing the time derivative of our desired quantities allows us to iteratively solve for the trajectories of the prisms:

$$r(t_i) = r(t_{i-1}) - \frac{1}{\eta_r}\frac{\partial U}{\partial r}\Delta t \tag{4.1}$$

$$\theta_I(t_i) = \theta_I(t_{i-1}) - \frac{1}{\eta_{\theta_I}}\frac{\partial U}{\partial \theta_I}\Delta t \tag{4.2}$$

where $i, i-1$ correspond to the $i^{th}, (i-1)^{th}$ time-step, respectively, and $I = 1,2$ corresponds to the particle.

The partial derivatives are taken of an interpolated interaction potential using the potential values determined via Surface Evolver on the regular $(r, \theta_1, \theta_2)$ grid, as discussed previously.

The viscous damping coefficients are not independent constants. They both originate from the interaction between the particle and the surrounding fluid. The center of mass drag $\eta_r$ depends on the particle orientation and the direction of center-of-mass motion. To our knowledge there is no literature on the fluid drag of triangular prisms, so in this study we make a simplifying assumption that both $\eta_r$ and $\eta_\theta$ are constants, and we estimate their magnitude by considering the following calculation: The work done over a small linear translation of $\Delta r$ due to the drag force is $W_l = F_d \Delta r$, while the work done over a small rotation by $\Delta\theta$ (in radians) due to the drag torque is given by $W_r = \tau_d \Delta\theta$. We can attribute the work done by each drag to the energy required to move the fluid due to the particle's motion. If we keep the small distance traversed by a single tip of the (equilateral) triangle the same in both cases, $\Delta r$, then the amount of rotation associated with that movement is given by $\Delta\theta = \Delta r/c$, where $c$ is the distance from the centroid to the tip. If the equilateral triangle has a side length of $s$, then $c = s/\sqrt{3}$. Comparing these two cases, the amount of fluid that is moved is of the same order, which means that we can equate $W_l$ and $W_r$. We also assume that these two motions require the same amount of time, $\Delta t$. In this case, we obtain

$$\eta_r \frac{\Delta r}{\Delta t} \Delta r = \eta_\theta \frac{\Delta\theta}{\Delta t} \Delta\theta, \tag{5}$$

so that the ratio between the two drag coefficients becomes

$$\frac{\eta_\theta}{\eta_r} = \left(\frac{\Delta r}{\Delta\theta}\right)^2 = c^2. \tag{6}$$

For angles measured in degrees, this equation becomes

$$\frac{\eta_{\tilde{\theta}}}{\eta_r} = \left(\frac{\Delta r}{\Delta\tilde{\theta}}\right)^2 = c^2 \left(\frac{\pi}{180}\right)^2. \tag{7}$$

For $s = 120~\mu m$, $c = 69.3~\mu m$ and $\eta_{\tilde{\theta}}/\eta_r = 1.46~(\mu m/°)^2$.

*Theoretical power-law relation for dilute binding trajectories*

For an experimental system exhibiting pairwise binding due to capillary interactions, the resultant trajectory can be characterized by the form of the separation distance $r$ as a function of time-to-contact, $t_c - t$, where $t_c$ is the first instance where the particles touch. If the trajectory obeys a power-law relation such that $r \sim (t_c - t)^\beta$, the exponent $\beta$ gives insight into the order of the capillary interaction, as we presently show.

The capillary interaction energy between two ideal multipoles is $U_{12} \sim r^{-\alpha}$, where $\alpha = 2m$ for an interaction between two multipoles of order $m$. Equating the resultant capillary force (for fixed orientations) to the viscous drag force yields a simple first-order differential equation

$$\frac{dr}{dt} \sim r^{-(\alpha+1)}, \tag{8}$$

which can be solved to obtain the desired result that the pairwise binding trajectory between two ideal capillary multipoles of order $m$ is characterized by a power-law exponent

$$\beta = \frac{1}{\alpha + 2} = \frac{1}{2(m+1)}. \tag{8}$$

Exponents of particular importance in this context are $\beta = 1/6$ (two quadrupoles) and $\beta = 1/8$ (two hexapoles).

**Results**

*Capillary-driven binding of triangular prisms at a flat air-water interface*

Prisms of all T/L ratios undergo lateral capillary-driven binding at a flat air-water interface. Capillary attractions yield particle-particle binding immediately upon particle attachment at the interface. Over a period of about one hour, the prisms self-assemble into open structures of progressively increasing size (as shown for the case of T/L = 1/25 in Figure ).

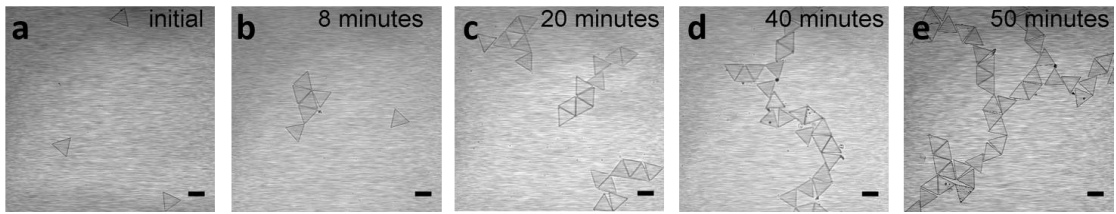

*Figure 3 Optical microscopy time-series images of capillary-driven triangular prism (T/L ~ 1/25) binding at a flat air-water interface. a) Initial placement of prisms at interface b) 8 minutes after placement of prisms at interface c) 20 minutes d) 40 minutes e) 50 minutes. Scale bars are 100 μm.*

*Polarity in interface attachment for thin prisms*

Figure shows 1270 x 1270 μm regions of open networks formed by prisms of the four T/L ratios synthesized. Each row in Figure corresponds to a specific T/L ratio (row 1 shows an open network formed by T/L = 1/50 prisms, row 2 is for T/L = 1/25 prisms, row 3 is T/L = 1/10, and row 4 is T/L = 1/5). The networks span several millimeters in space and are visible to the eye. For the three thinnest T/L ratios, the network's steady-state microstructure is comprised of a mix of dense, close-packed regions (with numerous prisms bound edge-to-edge), long strands, and large voids. On the other hand, relative to the thinner prisms, the network self-assembled from T/L = 1/5 prisms contains significantly fewer prisms in close-packing configurations, less chaining, smaller voids, and a generally more homogeneous prism density throughout the image.

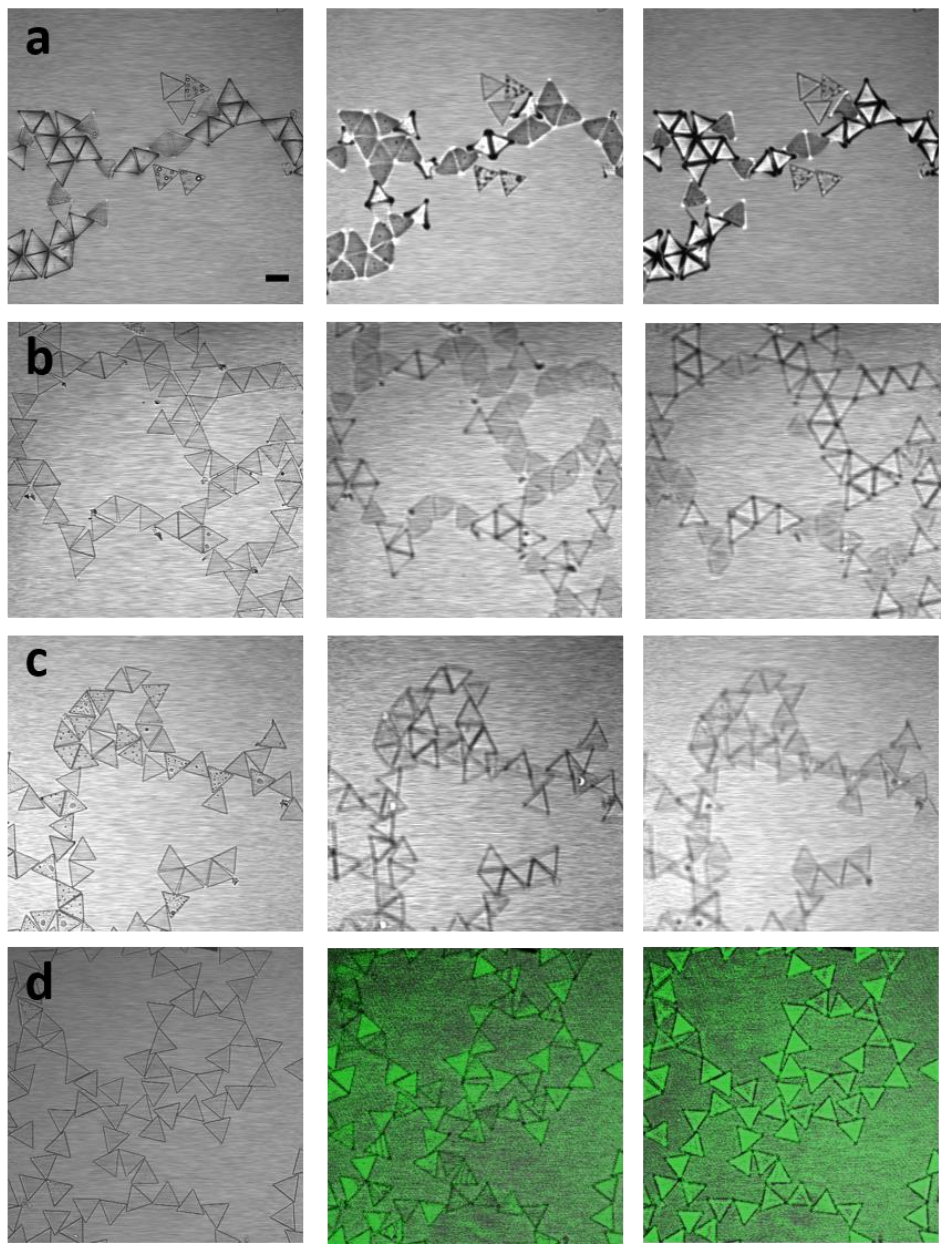

*Figure 4 Optical microscopy images of 1270 x 1270 $\mu m^2$ regions of open networks. Networks are self-assembled via capillary-driven triangular prism binding. Row 1 (a) T/L ~ 1/50, row 2 (b) T/L ~ 1/25, row 3 (c) T/L ~ 1/10, row 4 (d) T/L ~ 1/5. Column 1: single frame image of portion of network (1270 x 1270 $\mu m$), focal plane at air-water interface. Column 2: same single frame image of portion of network as in column 1, focal plane ~200 $\mu m$ below air-water interface. Column 3: same single frame image of portion of network as in columns 1 and 2, focal plane ~200 $\mu m$ above air-water interface. Green images in row (d) are overlays of optical and reflection microscopy; the reflection channel highlights differences in position of thick, apolar prisms at the flat air-water interface. Scale bar is 100 $\mu m$.*

In the course of imaging the open networks (c.f Figure 4), the location of the microscope's focal plane relative to the air-water interface was varied and an interesting feature of the pair binding was observed. Upon varying the focal plane slightly above and below the interface, we observe that T/L = 1/50, 1/25, and 1/10 prisms are pinned to the interface in such a way that their centers-of-mass either sit slightly above or below the interface. The second and third columns of Fig. 4 show this kind of imaging in the same 1270 x 1270 μm region of the open network as imaged in the first column. In column one, the microscope's focal plane is located at the air-water interface. All prisms are clear and visible, as demonstrated by their sharp, dark edges and tips, as well as their bright bodies. In the second column, the microscope's focal plane is located ~ 200 μm below the air-water interface. For the three thinnest prisms (T/L = 1/50, 1/25 and 1/10), some prisms remain clearly visible, with their dark edges and tips appearing thicker and even more discernable than in the first column and their bodies remaining bright, while all other prisms fall distinctly less visible, with their tips becoming bright and their edges and bodies appearing darker and faded.

In column three, the microscope's focal plane is located a similar amount *above* the air-water interface, in the opposite direction of the second column images. For the three thinnest prisms (T/L = 1/50, 1/25, and 1/10), the prisms that were clearly visible in the second column now appear faded, while the particles that appeared faded in column two are now clearly visible. (Additionally, a very small fraction of T/L = 1/50 prisms appear equally visible on both edges of the interface.)

This visual contrast in particles of opposite polarity is a scattering effect owing to transmission imaging, and indicates that the particles either sit below or above the interface. In the ensuing discussion, we define this as the "polarity" of the interface attachment. A prism with positive polarity refers to a prism whose center of mass sits above the interface in the assembly experiments, while a prism with negative polarity refers to a prism whose center of mass sits below the interface.

Although the three thinnest prisms are divided into populations located above and below the interface, the thickest prisms (T/L = 1/5) do not exhibit such visible vs. faded polarity; these prisms all appear equally visible relative to one another in both columns two and three. The relative image quality for the T/L = 1/5 prisms appears better below the interface (column two) than above (column three), suggesting that all these prisms are situated slightly below the interface.

To further investigate the precise manner of prism interface attachment, we observe the prisms using eSEM (Figure 5). Figure 5 confirms polarity in interface attachment for T/L = 1/50, 1/25, 1/10 but not for the thickest (T/L = 1/5) prism, consistent with the results from changing the optical microscopy focal plane. eSEM images of T/L = 1/50 prisms are shown in *Figure* a and d. *Figure* a shows a prism whose center of mass lies above the gelled interface in the water phase, and *Figure* d shows a prism whose center-of-mass lies below the gelled interface in the air phase. In addition, significant particle bowing along each of the three prism edges is observed.

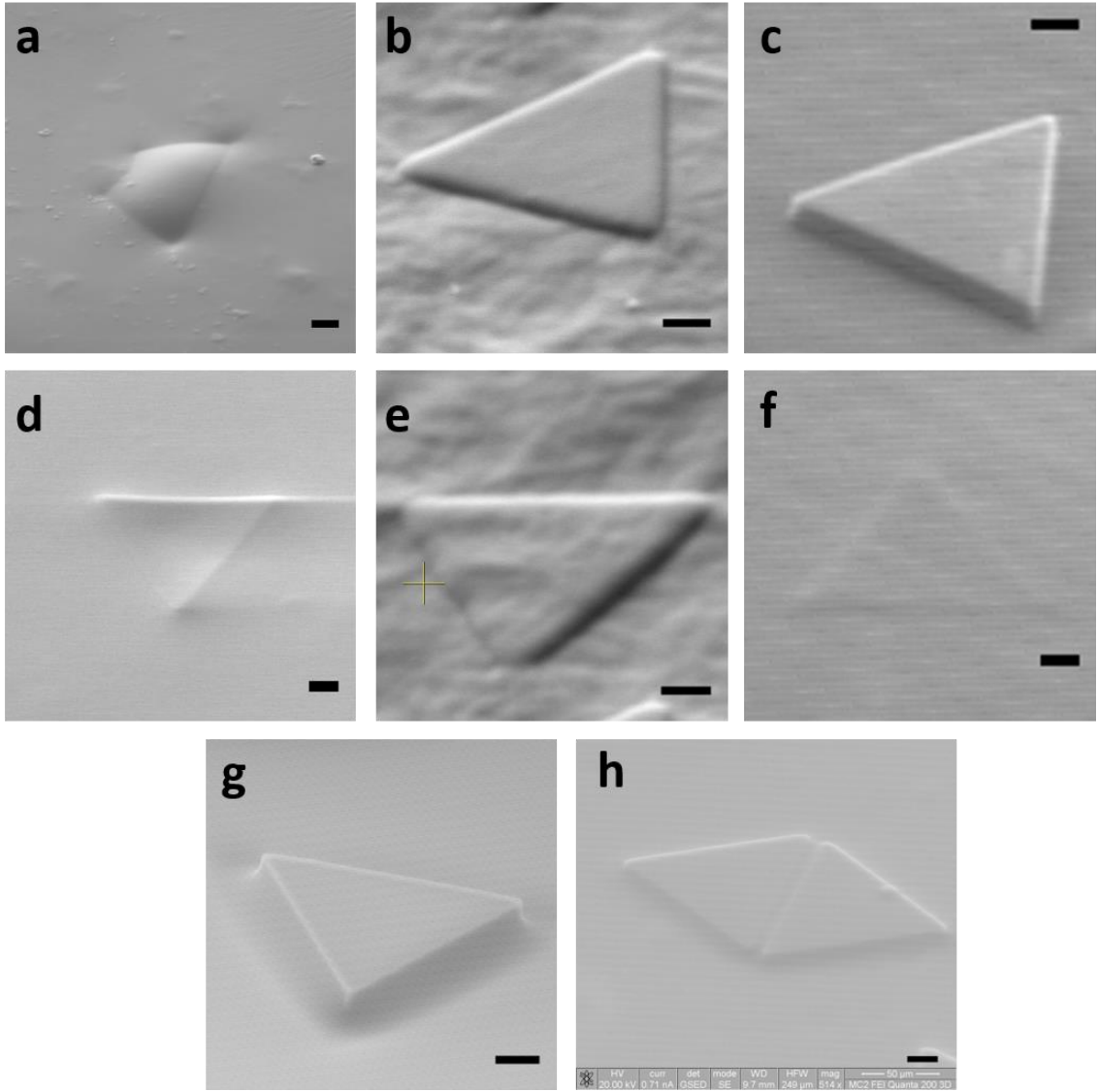

*Figure 5 Environmental SEM images of triangular prisms, fixed at an air-gellan/water interface. Row 1: (a) - (c) prisms assigned positive polarity: (a) T/L = 1/50, (b) T/L = 1/25, (c) T/L = 1/10. Row 2: (d) - (f) prisms assigned negative polarity: (d) T/L = 1/50, (e) T/L = 1/25, (f) T/L = 1/10. (g) apolar T/L = 1/5 prism. (h) The same capillary-driven binding states are observed at air-gellan/water interface prior to prism immobilization as are observed with optical microscopy at non-gelled interfaces. Scale bars are 20 μm.*

Interface attachment of the T/L = 1/25 prisms are shown in *Figure* b and e. The top face of a prism in *Figure* e is covered by the gelled interface (as evidenced by the rippling texture on top of this prism, which is consistent with the surface of the gelled water phase elsewhere in the image), while several other prisms in the image sit with their top faces uncovered by the interface (as evidenced by the smooth texture of the exposed faces of these particles, relative to the rippling surface of the gelled water phase). The T/L = 1/25 prisms do not appear as bowed as the T/L = 1/50 prisms. Still, evidence for polarity in prism-interface attachment is apparent because the covered prisms' centers of mass sit below the interface (in the gelled water phase), and the uncovered prisms' centers of mass sit above the interface (in the air phase). Polarity is again observed for T/L = 1/10 prisms, shown in *Figure* c and f. Several prisms rest with their centers of mass below the interface, and the top face of the prism is covered by the surface of the gelled interface, while other prisms sit substantially higher on the interface, with their top faces exposed to the air phase. Polarity of the particle position relative to the interface is not apparent for T/L = 1/5 prisms. Fig. 5g is representative of all observed T/L = 1/5 prisms; the interface is observed to rise at the corners of the prisms, and prisms all appear to sit at the same interface position, relative to both the interface and to one another.

The optical micrographs also show evidence for bowing in T/L = 1/50 (Fig. 5a) and 1/25 (Fig. 5b) prisms. That is, prisms of assigned polarity appear to have bright, central bodies and dark tips when particles reside on the same side of the interface as the focal plane, and faded central bodies and bright tips when they reside on the opposite side of the interface as the particles. This illumination contrast appears consistent with a difference in the position of the prism central body and tips relative to the microscope's focal plane. Moreover, referring back to Figure , bowing is apparent in the SEM images of the particles as originally fabricated. Apparently, this bowing is a permanent, reproducible feature of the thin prism fabrication; it persists from synthesis to assembly.

Bowing can specify the curvature of the interface at the prism boundary by contact line pinning. This interfacial curvature in turn determines the capillary-driven attraction between the prisms. *Figure* show that prisms with positive polarity (on top of the interface) are bowed downwards (with tips pointing towards the water phase), and negative polarity (below the interface) prisms are bowed upwards (with tips pointing towards the air phase). In both cases, the interface appears pinned to the corner of the prism's edge and to the concave face. Thus, the curvature of the interface follows the curvature of the bowed prism. The result is that the interface curvature at the tips and edges of triangle is opposite for prisms of positive and negative polarity.

Inhomogeneity in prism surface wetting – which could potentially be introduced during prism fabrication as described in the methods – is not the source of prism polarity. Thin prisms (T/L $\leq$ 1/10) fabricated with or without plasma treatment on one side each exhibit the two polarity states. The plasma treatment affects wetting; the contact angle change in the plasma treated particles is ~70° immediately following treatment. This insensitivity to plasma treatment supports the hypothesis that prism bowing is the primary driver of the observed polarity.

*Correlation between prism interface polarity and bonding state*

The correlation between particle polarity (up or down interface attachment) and bonding state is examined for T/L = 1/50 prisms in Figure ; Comparable measurements for T/L = 1/25 are available in SI Figure 3. Each row of Figure shows one 1270 x 1270 µm region of an open network. In the first column, the focal-plane is located above the interface. In the second column, the focal-plane is located below the interface. In these first two columns, each prism is assigned a polarity, determined by the location of the prism center-of-mass, as described in the previous section. For T/L = 1/25, a polarity is assignable to all prisms.

In the third and fourth column, the focal-plane is located at the interface. In the third column, bonds between prisms with (a) the same polarity (identified with red and blue connecting lines for bonds between pair-bonded prisms of negative and positive polarity, respectively) (b) the opposite polarity (purple connecting lines), and (c) indeterminate polarity (black connecting lines) are predicted. In the fourth column, prism-prism bonds are measured by the relative orientation of the two particles, independent of the polarity state of each particle. Four types of bonds are observed: (a) tip-tip (green connecting lines), (b) tip-edge (pink connecting lines), (c) edge-edge (orange connecting lines; edges of triangles are in registry – in contact and flush with one another), and (d) edge-edge offset (brown connecting lines; half of the edge of each bonded triangles lie flush with one another, with the tip of one triangle located at the center of the other triangle's edge).

Comparison of the predicted and measured bonded states for $T/L = 1/25$ and $T/L = 1/50$ prisms shows perfect correlation between the polarity states of any two adjacent particles and their bonded state. Specifically, of all prisms whose polarity could be determined, all bonding between same polarity prisms is tip-tip or edge-edge, and all bonding between opposite polarity prisms is tip-edge or edge-edge offset. The bonded states – both measured and predicted based on polarity – are available in SI Tables 1 ($T/L = 1/25$) and 2 ($T/L = 1/50$).

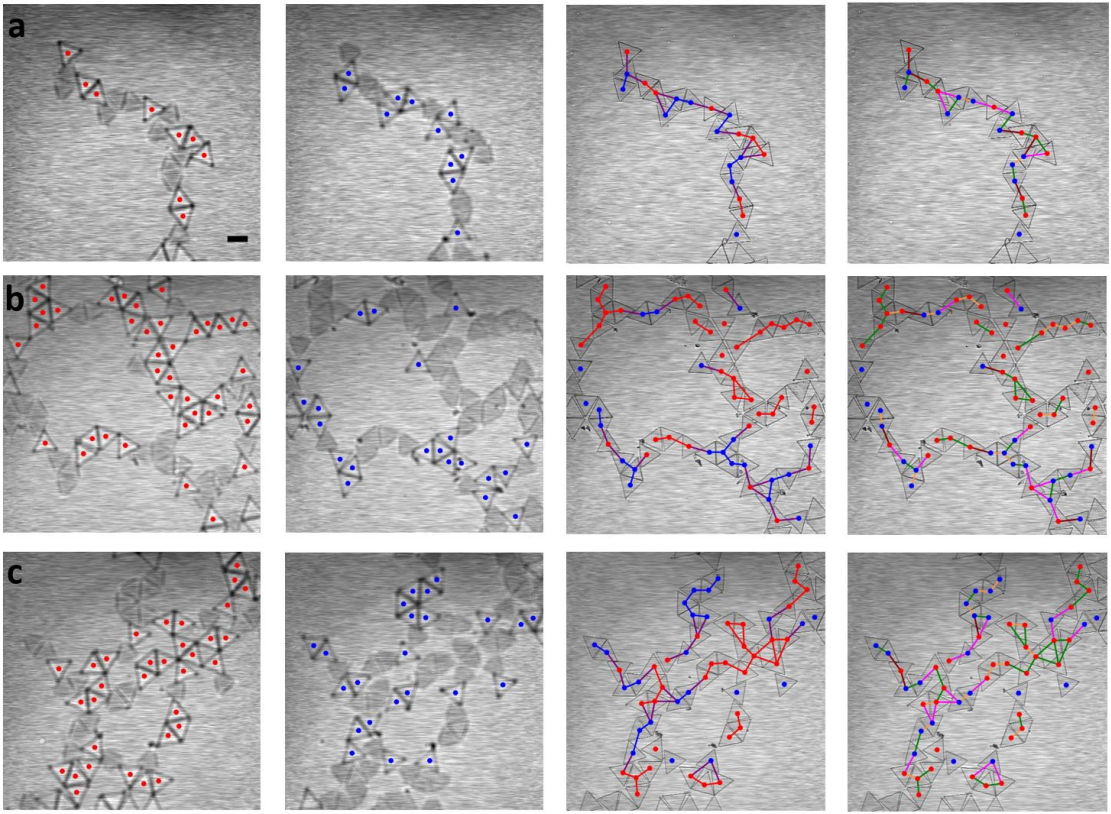

*Figure 6 Identification of triangular prism binding states (T/L = 1/25). Each row of images (a) – (c) represents a different location within a network structure. The relative position of microscope's focal plane to the air-water interface is varied by column as follows: Column (1): Microscope focal plane is ~200 μm below the interface. Clearly visible prisms are identified with red markers. Column (2): Microscope focal plane is ~200 μm above the interface. Clearly visible prisms are identified with blue markers. Column (3): Microscope focal plane is at the interface. Bonds between prisms with the same polarity are identified with blue and red connecting lines, bonds between prisms with the opposite polarity are identified with purple connecting lines. Column (4): Microscope focal plane is at the interface. Prism-prism bonds are identified by their polarity-independent orientation: side-side (orange connecting lines), tip-tip (green connecting lines), side-side offset (brown connecting lines), tip-side (pink connecting lines). Bonds in Columns (3) and (4) are tabulated in Table (S1). Scale-bar is 100 μm.*

*Theoretical analysis and computation of capillary interactions of triangular prisms*

The triangular prisms in these experiments have flat, nearly vertical side surfaces. This lack of curvature of the particle sides leads to a different kind of interface attachment than that observed with ellipsoids and cylinders. As discussed in [8–10,37], interfaces around the ellipsoids and cylinders either rise or fall as a result of variations in curvature of the side surface of the particle. A constant contact angle as well as zero total force and torque on an isolated particle is maintained.

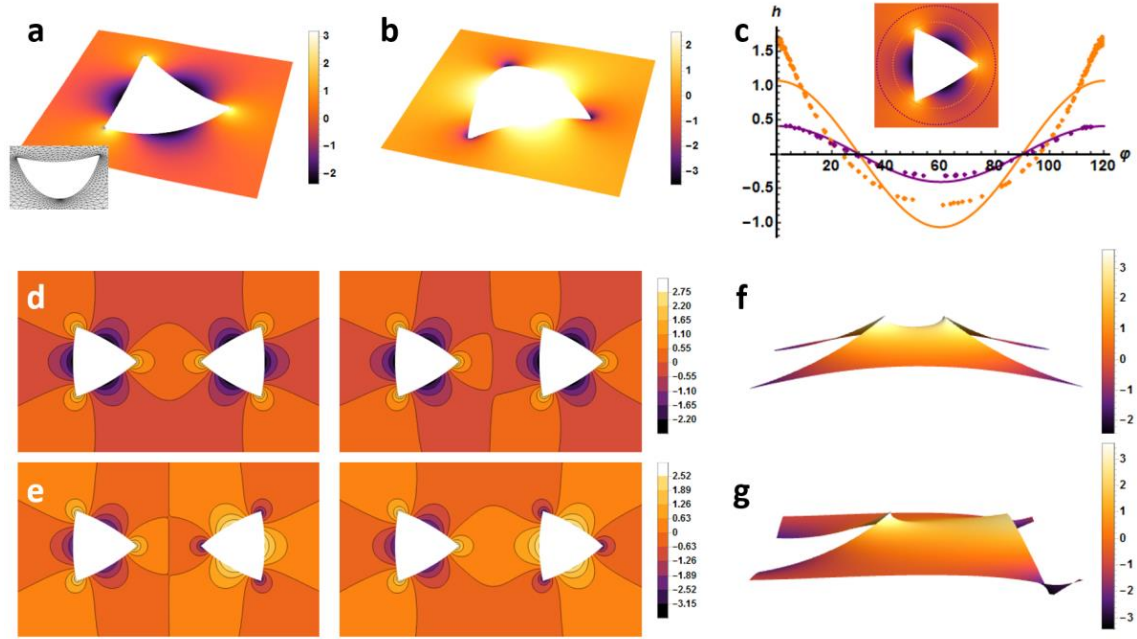

*Figure 7 Interface height profile for a (a) negative polarity bowed-up triangular prism and a (b) positive polarity bowed-down triangular prism, where the zero value is set by the equilibrium interface height at large distances from the prism. The inset in (a) is a close-up of the Surface Evolver simulation output. (c) A comparison of the interface height profile around a bowed-up triangular prism (data points) and an ideal hexapole (solid curves) as a function of angle at two different distances from the triangular prism, shown in the inset. Simulated interface height profiles for (d) two bowed-up triangular prisms and (e) one bowed-up and one bowed-down prism for both tip-to-tip and tip-to-side configurations. Zoomed-in rendering of simulated interface height profile for (f) a tip-to-tip configuration for two bowed-up prisms; and (g) a tip-to-side configuration for one bowed-up and one bowed-down prisms, illustrating the existence of a capillary bridge in both cases.*

For these triangular prisms with vertical side surfaces, however, the preferred contact angle (~ 5 degrees) of the material cannot be reached, because it would correspond to a uniform rise of the interface around the triangular prism, yielding a net force pointing down on the particle; this net force is inconsistent with mechanical equilibrium. Therefore, instead of an equilibrium contact line in the middle of the side surface of the triangular prisms, the interface is pinned to the edges of the concave face of the triangular prisms with a non-equilibrium contact angle that satisfies mechanical equilibrium (see Supplementary Information for further discussion of this phenomenon). Contact-line pinning has been observed in various experimental systems consisting of solid particles or substrates containing sharp edges.[38–41]

To characterize the interface shape and the resulting capillary interaction between the triangular prisms, we use Surface Evolver to compute the interface with a contact line pinned to the edges of a bowed triangle (for details see the Materials & Methods section). To match the observed curvature of the thinnest particles, we use an inverse-curvature-to-edge-length ratio of 0.9 (that is, for an edge length of 120 $\mu$m, we take the radius of curvature to be 108 $\mu$m). The resulting interface around isolated particles (Figurea,b) closely resembles that observed in the eSEM images of the thinnest particles (*Figure* ).

It is worth noting that the only input into the Surface Evolver computation is the pinned contact line, and no information about the particle thickness is involved. Our computation shows that, for a bowed-up particle (Figurea), the particle center of mass is below the interface in the far field by 7.45 µm (for the given curvature), whereas the center of mass of a bowed-down particle is the same amount above the far-field interface. This depth is greater than the particle thickness, and explains the perfect correlation between the polarity and the direction of the particle bowing of the thinnest particles. (The relation between the interface attachment and the bowing direction of the thicker particles may involve mechanisms such as variability in interface height due to roughness and interface pinning; these mechanisms lead to weak quadrupolar interactions, as described in ref.[42 42]).

The interface geometry around the triangular prisms is similar to that of the capillary hexapole (discussed in detail in Supplementary Information) in that there are six regions of alternating positive- and negative- interface heights (where the equilibrium, unperturbed height of the interface at far distances is taken to be zero). However, important differences exist between the ideal hexapole field and the interface around the triangular prisms at distances close to the particle. The ideal hexapole field with height $h \sim \frac{1}{r^3} \cos 3\theta$, has the symmetry that the positive and negative regions are of equal width. The interface around the triangular prisms, in contrast, has much narrower positive (negative) regions around the tip of the bowed-up (-down) triangles (Figurec). As a result, the focusing of excess area around the tips of the triangles induces stronger capillary interactions at the tips than along the triangle edges. Note that, as one would expect, the height of the interface around a bowed triangular prism increasingly conforms to the profile of a capillary hexapole as the distance from the prism increases. Indeed, the effect of tips, edges, and other sharp particle features, which are quite prominent in the near-field behavior of the interface, becomes increasingly diminished and smoothed out at these larger distances (Figurec).

We study the capillary interaction potential between triangular prisms by computing the interface geometry around a pair of triangular prisms using Surface Evolver. Once the numerical interface solution has been obtained, we can subsequently determine the capillary interaction energy using $U_{12} = \gamma(\delta S_{12} - \delta S_1 - \delta S_2)$, where $\gamma$ is the air-water surface tension, $\delta S_{12}$ is the excess area created at the interface in the full two-particle system, and $\delta S_i$ ($i = 1,2$) is the excess area in an isolated one-particle system (i.e., for separation distance $r \to \infty$). The excess area is defined as the difference between the actual surface area $\Sigma^*$ and the projected surface area $\Sigma$ (see Supplementary Information). There are, of course, two cases to be simulated: the first is when both prisms have the same bowing polarity (by symmetry, we need only consider the case where both particles are bowed up), and the second is when the two particles have opposite polarities (here again we can simplify matters and consider only the case where the particle on the left is bowed up and the particle on the right is bowed down). Examples of the interface in the vicinity of two triangular prisms with the same and opposite polarities are shown in Figured,e.

It is already evident from these plots – even before further analysis – that the tip-tip configuration for prisms with the same polarity and the tip-edge configuration for prisms with opposite polarities are attractive, while the opposite configurations are repulsive – the former will result in decreased excess area as the particles move towards each other, while the latter will result in increased excess area (the overall slope of the interface will increase between the bowed-up and bowed-down components as they are brought closer together). Figuref,g show the underlying mechanism that reduces the excess area between regions with the same capillary charge: the formation of a capillary bridge.

The capillary interaction potential $U$ depends on both the distance between the centers of the two triangular prisms, $r$ and their orientations relative to the line connecting their centers, $\theta_1, \theta_2$ (Figure  ). This is a configuration space that has one extra dimension beyond that of the capillary hexapolar theory discussed in Supplementary Information, in which only the relative orientation of the two particles matters. In order to be able to directly compare the theoretical case with that of two bowed-up triangular prisms, we fix the orientation of the left particle to be 0° and allow $r$ and $\theta_2$ to vary. The resultant potential, shown in Figure a, is very similar to that of the ideal hexapoles; even the general shapes of the interfaces, as shown in a few select cases as insets in both plots, share similar features.

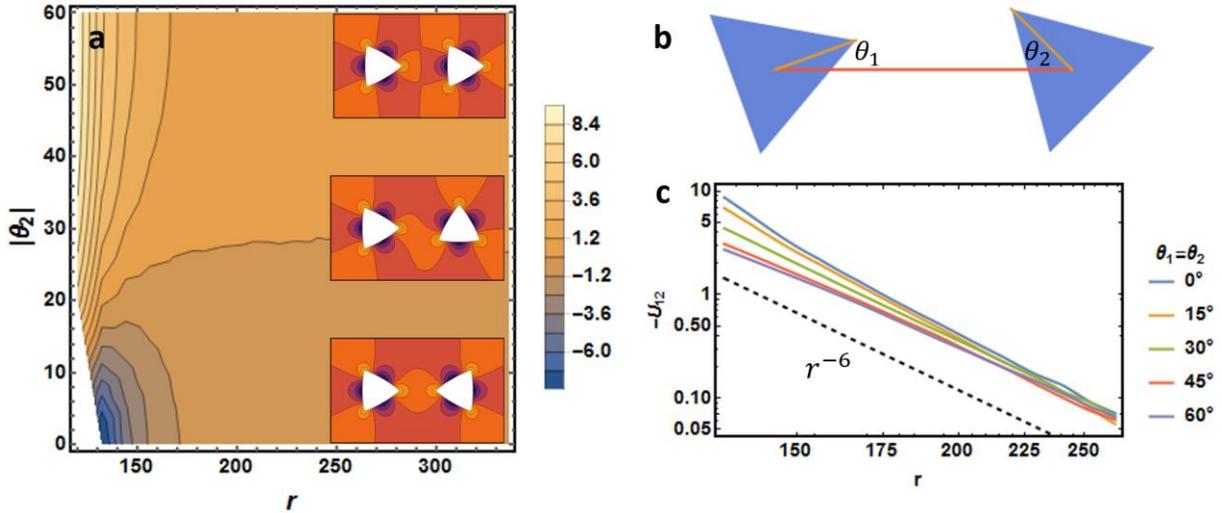

Figure 8 Numerically-simulated capillary interaction potential between two bowed-up triangular prisms, with the left prism held at 0°. This two-dimensional slice of the full three-dimensional configuration space is directly comparable to the theoretical interaction potential in Figure S2. (b) All orientation angles for the triangular prism system are defined according to the convention shown: the orientations are defined by the angle a specific tip of the prism makes with the line connecting the centers of the two prisms. (c) The capillary interaction potential for two-bowed up triangular prisms in mirror-symmetric configurations as a function of the separation distance, $r$, on a log scale, for various orientation angle values. A dashed reference line, corresponding to the theoretical interaction potential for two ideal hexapoles, $U \sim r^{-6}$, is shown for comparison.

The similarities extend beyond this, as well: in Figure c, when comparing potential curves for various mirror-symmetric configurations in the triangular-prisms system with that of the mirror-symmetric curve in the ideal-hexapoles system, which has a $r^{-6}$ dependence, we see that all the curves approach the theoretical ideal-hexapole curve at long distances, as we expect from the interface profile. Deviations from the ideal-hexapole curve and from each other occur at short inter-particle distances, where the anisotropic tips become increasingly prominent. Note that the $\theta_1 = \theta_2 = 0°$ tip-tip mirror symmetric configuration is favored for these smaller distances.

This deviation from an ideal hexapole is further portrayed in Figure . In the case of ideal hexapoles, since the interaction energy depends only on the relative orientations of the

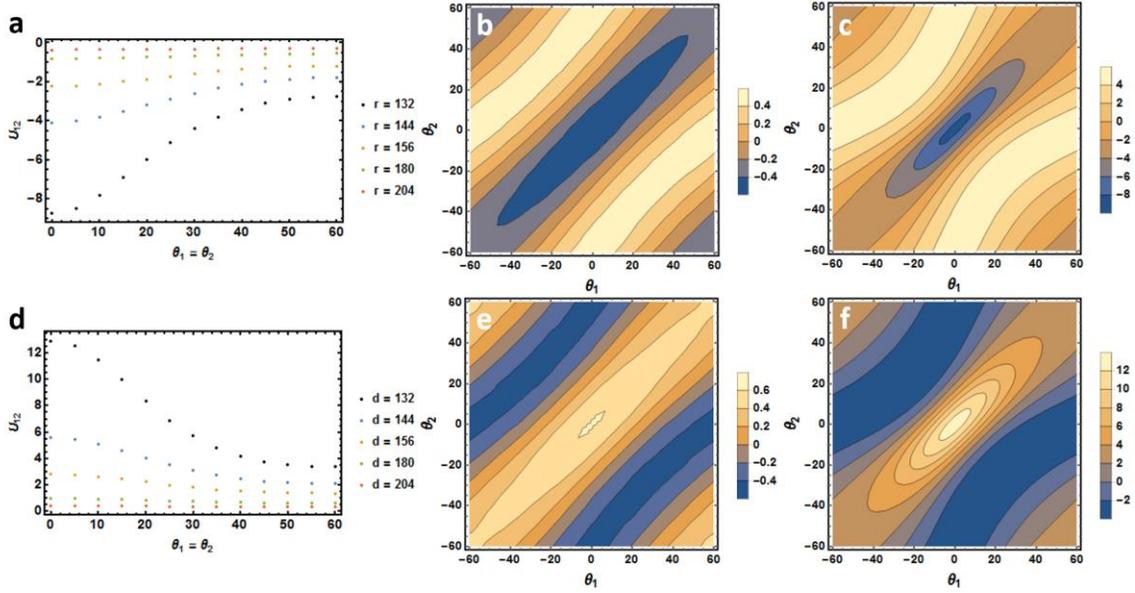

*Figure 9 (a) Interaction energy potential values for two bowed-up triangular prisms in mirror-symmetric configurations at various separation distances. 0° corresponds to a tip-to-tip configuration, while 60° corresponds to a side-to-side configuration. Interaction energy potentials plotted as a function of orientation angles for (b) $r =$ 192 μm and (c) $r =$ 132 μm. (d)-(f) The corresponding three figures for the case of one bowed-up and one bowed-down triangular prism.*

particles, the interaction energy for all mirror symmetric configurations, in which $\theta_1 = \theta_2$, for a given distance is perfectly degenerate. As shown in Figure a,d, for the system of triangular prisms, however, the tip-tip mirror symmetric configuration (corresponding to $\theta_1 = \theta_2 = 0°$) is strongly favored (disfavored) compared to the edge-edge mirror symmetric configuration (corresponding to $\theta_1 = \theta_2 = 60°$) for smaller values of inter-particle distances, $r$ in the same- (opposite-) polarities system. In the case of opposite polarities, even though the edge-edge configuration is preferred over the tip-tip configuration, it is important to realize that it is *not* the global preferred state, which is a non-mirror-symmetric configuration, as will be discussed further subsequently. Once again, in both cases, the expected ideal-hexapole behavior of degenerate energies for all mirror symmetric configurations is recovered as the inter-particle distance is increased.

As shown in Figure b,c, the potential for a pair of bowed-up triangular prisms shows a clear well for the mirror symmetric configuration, $\theta_1 = \theta_2$, which becomes increasingly deep for smaller inter-particle distances. It is clear in Figure b, as well, that for two bowed-up triangular prisms, the potential is relatively flat for all mirror symmetric configurations at a given large distance (same as ideal hexapole interaction). However, when the two prisms are close, the tip-to-tip configuration is much more preferred (in contrast to the ideal hexapole). The above results indicate that when two bowed up particles approach one another, in general, they first rotate into mirror symmetric configurations, and then rotate to tip-to-tip when they are very close to each other. The case of two bowed-down triangular prisms is very similar to the discussion above for the bowed-up case, with the simple addition of a minus sign of the interface height, which results in the same interface energy.

The case of one bowed-up triangular prism and one bowed-down triangular prism is quite different. At large distances, the capillary interaction is close to that between two hexapoles but with one hexapole rotated by 60° degrees (or equivalently the "+" and "−" capillary charges switched). Interestingly, at small distances, the potential energy valley appears curved in $(\theta_1, \theta_2)$ space while slightly favoring offset edge-edge configuration (Figure e). As we see below, this leads to different binding trajectories for bowed- up pairs and up-down pairs.

*Dilute binding events: experimental observations*

In order to evaluate the modeling of the capillary interaction between the triangular prisms, we simulate pair trajectories of prisms from various initial conditions, and compare these trajectories with trajectories observed in experiments. The centroidal separations, r, and angular orientations of the two prisms relative $\theta_1$ and $\theta_2$ were collected by image analysis. The trajectories are available in Movies S1 through S7. SI Figure 4a,b,c, and d show frames from the trajectories from Movies S1, S2, S4 and S7, respectively.

Seven trajectories (five for T/L = 1/25 and two for T/L = 1/50) were collected from the SI movies; separations and orientations are reported in Figures 10 and 11. Four of the trajectories report like polarity binding (c.f. Fig. 10). Three trajectories report opposite polarity binding (c.f. Fig. 11). Qualitative features observed for dilute binding trajectories are: (i) like polarity particles, in a first stage adopt mirror symmetric configurations and slowly move toward each other; in a second stage, particles rapidly close into a tip-tip binding; and in a third stage, some particles then rotate into a edge-edge configuration; (ii) opposite polarity particles approach to a tip-midpoint edge configuration; the pair finally collapses into an offset edge-edge bond.

Figure also shows that the time for binding of T/L = 1/50 prisms is significantly faster than for T/L = 1/25 prisms. This difference indicates that capillary attractions are much stronger at separation distances of up to several prism edge lengths for T/L = 1/50 prisms as compared to T/L = 1/25 prisms. By contrast, there is negligible difference in the time scale of tip-tip and tip-edge binding at fixed T/L ratio, an indication that the strength of like-polarity and opposite-polarity interactions are similar. Furthermore, the exponent associated with particles approaching each other in these binding events, $\sim(t_c - t)^\beta$, where $t_c$ is the time of contact, defined as the first image frame where the two prisms touch, displays similarity with the exponent from hexapole-hexapole interactions, $\beta_0 = 1/8$ (Figure ). The small deviation comes from the difference between the actual capillary interactions between the triangles with the ideal hexapolar interaction. In particular, at far distances, $\beta$ appears to be closer to 1/6, indicating that at far-field quadrupolar interactions (from random variations in prism edge topography) may be the dominant driver for binding at these large separations. Nevertheless, the scaling at small separation approaches the hexapolar expectation of 1/8.

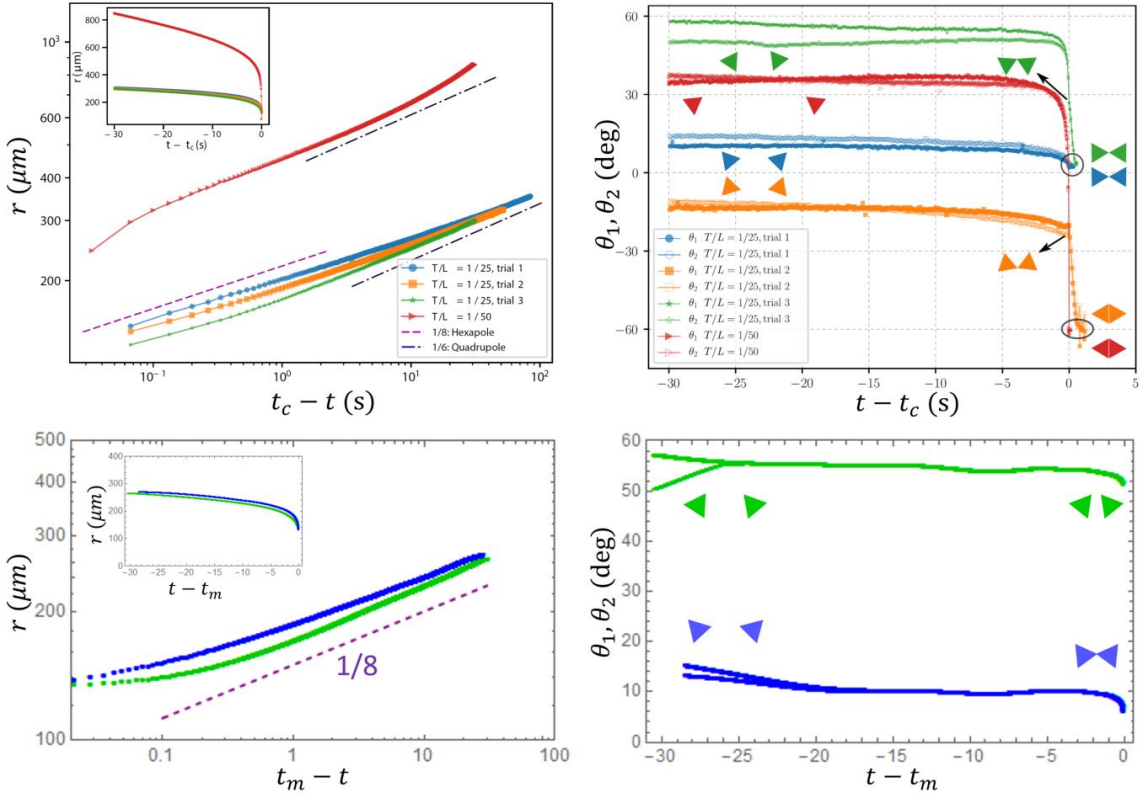

*Figure 10 Comparison of experimentally observed and simulated trajectories for a pair of prisms of same polarity. Top row: observed r vs $t_c - t$ curve in log-log scale (left) and linear scale (inset), where $t_c$ is taken to be the first frame in which the two prisms touch; and observed $\theta_1, \theta_2$ vs $t - t_c$ curves (right). Four events are shown as explained in the legend, and lines showing $\beta = 1/8$ (consistent with hexapolar interaction) and 1/6 (consistent with quadruplar interaction) are added. Illustrations of the prisms configurations are added in the $\theta_1, \theta_2$ plot to show the geometry. Configurations at the time of contact ($t = t_c$) are pointed to by arrows, and the points at $t - t_c > 0$ show prism rotations after contact, with final configurations marked by circles. Bottom row: counterparts of the r and $\theta_1, \theta_2$ plots from simulation. Instead of contact time, $t_m$ is the time where the prisms' separation distance reaches $r_m = 132$ μm (the lower bound of r in our computation), at which they touch if $\theta_1 = \theta_2 = 0$. We have chosen initial conditions that are close to two experimental trajectories.*

Turning to prism rotation, for tip-to-tip trajectories, prism rotation begins between hundredths of a second (T/L = 1/25) up to several seconds (T/L = 1/50) prior to contact (Fig. 10). In the later case, these times correspond to separation distances that are several edge lengths. The angular orientation plots also show that prisms bind in a mirror symmetric fashion; that is, in each pair-binding event, both prisms rotate an equal amount into their final, steady-state orientation. For tip-to-midpoint edge trajectories, prism angular orientation also begins at distances corresponding to separations of several edge lengths (Fig. 11).

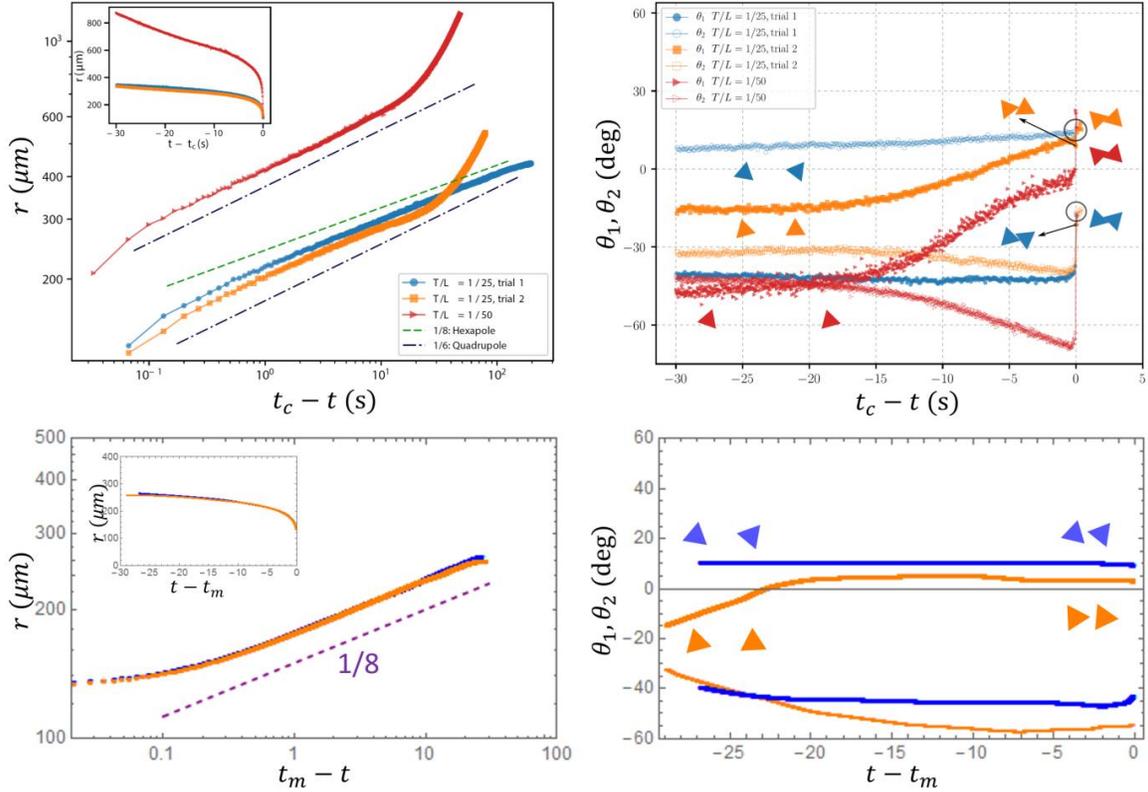

*Figure 11 Comparison of experimentally observed and simulated trajectories for a pair of prisms of opposite polarity. Top row: experimental observations. Bottom row: simulation results. All conventions are the same as in Figure . Note that in two experimental events, the prisms approach faster in the far-field regime than quadrupolar interactions would dictate; we speculate that this is due to noncapillary-induced drift of particles at the interface, possibly owing to convective flow at the interface surface .*

*Dilute binding events: simulation results and agreement with experiments*

To compare to these results, we simulated pair-binding events using the interaction potential (interface energy) $U(\theta_1, \theta_2, r)$ obtained by interpolating a grid of Surface Evolver-calculated potential values at regular intervals as described above. Details of the trajectories simulation are described in the Materials & Methods section.

Examples of our simulation results for same-polarity (both bowed up in our calculation) and opposite-polarity prisms are shown in Figures 12 and 13, respectively. Initial conditions were chosen to approximate trajectories observed experimentally. The ratio of the two viscous-damping coefficients is taken be to $\eta_{\tilde{\theta}}/\eta_r = 1.46$ as discussed in Materials & Methods, and $\eta_r$ is taken to rescale time such that the arbitrary time scale in the simulation approximates the experimental time scale, in units of seconds. These simulations terminate at $r_m = 132 \ \mu$m, the distance at which the two prisms would touch if they faced one another tip-to-tip. In all cases, the far-field trajectories are roughly consistent with that of the ideal hexapole-hexapole interaction, while expected, and important, deviations from the ideal interaction occur as the separation distance decreases. In Supplementary Information we show additional trajectories where additional initial conditions with different choices of $\eta_{\tilde{\theta}}/\eta_r$ ratio are discussed.

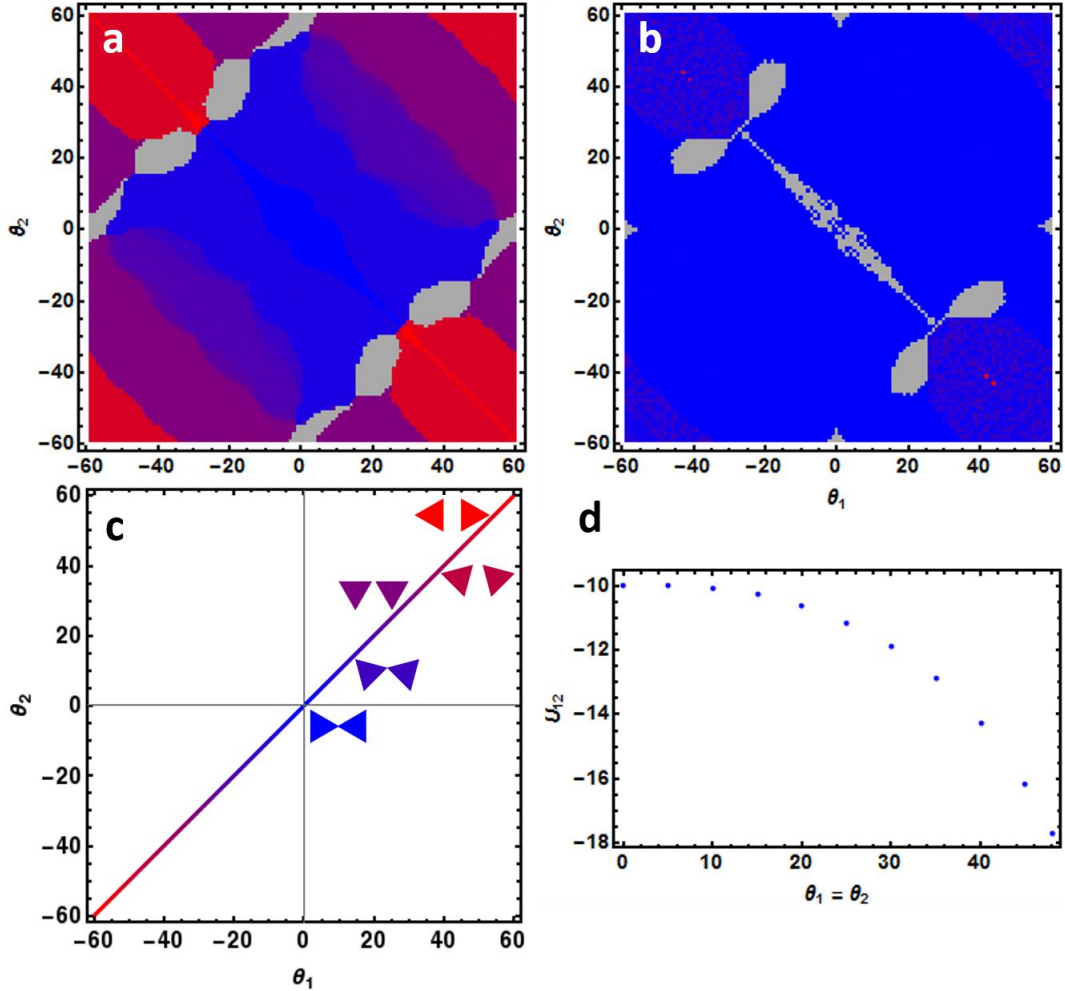

*Figure 12 "Phase" diagrams illustrating the final configurations ($r = 132$ μm) for all possible initial orientations for two bowed-up triangular prisms at $r = 264$ μm for two different viscous-damping coefficient ratios, (a) $\eta_{\tilde{\theta}}/\eta_r = 1.46$ (μm/°)$^2$ and (b) $\eta_{\tilde{\theta}}/\eta_r = 0.146$ (μm/°)$^2$. The final configurations are all mirror-symmetric and lie somewhere along the line in (c), with blue corresponding to tip-to-tip final configurations, red corresponding to side-to-side final configurations, and gray denote initial conditions that leads to trapped configurations which we believe to be artifacts of the computed pair potential.. (d) Capillary interaction potential values for mirror-symmetric configurations with two tips of the triangular prisms remaining in contact (thus, the separation distance, $r$, decreases below $132$ μm as $\theta_1 = \theta_2$ increases. Indicates a tendency for tip-to-tip configurations to ultimately collapse to side-to-side configurations.*

To obtain statistics about how the triangular prisms bind, we ran simulations for all initial angles of prism pairs at an initial distance of $r_0 = 264\ \mu m$; our results are summarized, for two different ratio values, $\eta_{\tilde{\theta}}/\eta_r = 1.46\ (\mu m/°)^2$ and $0.146\ (\mu m/°)^2$, in Figure 12a,b. The first ratio is chosen according to the simple geometric estimate discussed previously. The second ratio, which is 10 times smaller, allows the prisms to rotate faster relative to their center of mass motion, and presents a useful contrast to the first case. In both cases, a significant majority of configurations end up in, or close to, the $\theta_1 = \theta_2 = 0°$ tip-tip mirror-symmetric configuration. For the case of $\eta_{\tilde{\theta}}/\eta_r = 1.46\ (\mu m/°)^2$, some trajectories end up along a continuum of mirror symmetric configurations ranging from tip-to-tip to edge-to-edge, as the prisms did not have enough time to finish the rotation before contact. Contrastingly, in the second case with $\eta_{\tilde{\theta}}/\eta_r = 0.146\ (\mu m/°)^2$ almost all trajectories end up tip-to-tip, because rotational drag is smaller, leading to faster rotation. A small fraction of initial conditions (gray in the figure) ended up in random configurations coming from small kinks in the pair potential due to numerical error.

In order to investigate what happens after the two prisms touch at their tips, we calculated the pair potential for two prisms in which their tips continue to touch, but at different orientations (mirror symmetric configurations with $\theta_1 = \theta_2$ ranging between 0 and 60°), as shown in Figure 12d This indicates that, after initial tip-to-tip contact, the pair of triangular prisms will rotate and "collapse" into an edge-to-edge configuration. It is worth pointing out that although – after collision – the prisms collapse into the edge-to-edge configuration, a majority of trajectories still first go through an initial tip-tip binding. This is in good agreement with our experimental observations.

For bowed-up-bowed-down pairs, a similar set of simulations yields results shown in Figure a,b for the two viscous-damping coefficient ratios. In this case, it is important to note that the final configurations are not mirror symmetric; Figure c shows the final orientation values for the two triangular prisms. The curves of final orientation lie along the minimum-energy regions of the opposite-polarity interaction potential in Figure f.

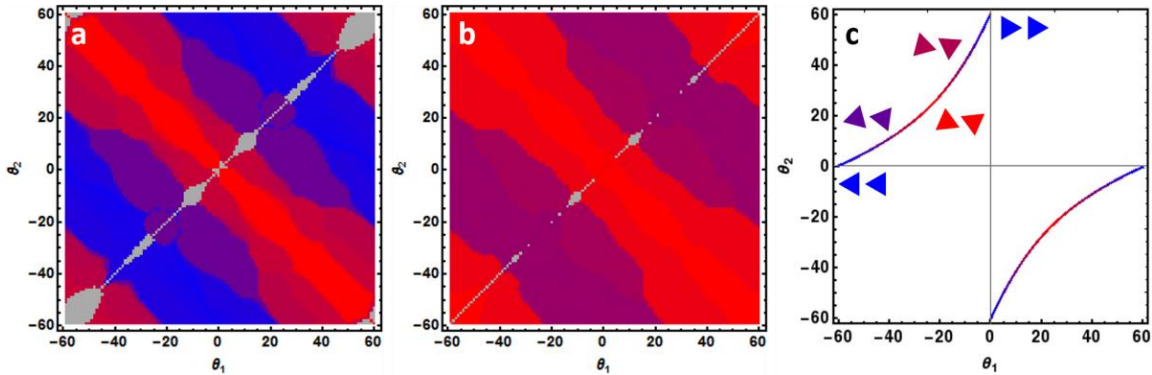

Figure 13 "Phase" diagrams illustrating the final configurations ($r = 132\ \mu m$) for all possible initial orientations for one bowed-up and one bowed-down triangular prism at $r = 264\ \mu m$ for two different viscous-damping coefficient ratios, , (a) $\eta_{\tilde{\theta}}/\eta_r = 1.46\ (\mu m/°)^2$ and (b) $\eta_{\tilde{\theta}}/\eta_r = 0.146\ (\mu m/°)^2$ . The final configurations lie somewhere along the curve in (c), with blue corresponding to tip-to-edge final configurations and red corresponding to offset-edge-to-edge final configurations.

Similar to the case of prisms with the same polarity, simulating the approach of prisms of opposite polarities with the smaller drag coefficient ratio, $\eta_{\tilde{\theta}}/\eta_r = 0.146\ (\mu m/°)^2$, leads to a more uniform state diagram wherein all initial conditions have enough time to rotate to the offset edge-to-edge configuration, which is of lower energy. The first case – which uses the ratio from our geometric estimation – yields a continuum of final configurations. As before, our simulation terminates at $r_m = 132 \ \mu m$, which is the distance at which two prisms touch when they face one another tip-to-tip. The opposite-polarity prisms, however, being in non-tip-to-tip configurations, are not yet touching at this distance. Our additional computations of the interface energy shows that, at smaller distances, the offset edge-to-edge configurations exhibit lower energy, leading to the final collapsed offset edge-to-edge configurations as observed in experiment.

*Assembly into open networks*

The pair-binding observations discussed above indicate that self-assembly of thin, triangular prisms may result in 2D networks with both open (tip-tip and tip-edge pair-binding orientations) and close-packed (edge-edge and edge-edge offset pair-binding orientations) conformation. We characterize statistical signatures of the resulting disordered networks for these prisms, which will guide future study aiming at obtaining regular open networks.

Recalling Fig. 3, at early times (Fig. 3b and c), small aggregates form. These small aggregates undergo time-dependent growth via aggregate-aggregate attraction and binding (Fig. 3d and e). These larger aggregates branch laterally, which yields an open structure. Aggregates continue to attract and bind to one another until all available prisms are incorporated into a space-spanning, open network. Representative networks formed by self-assembly are shown for all T/L ratios in Fig. 14. Each of the four images in Fig. 14 is a 3.8 x 2.5 mm spatial mosaic of either six or eight (either three-by-two or four-by-two) 1270 x 1270 μm² microscopy images. While each of the self-assembled networks possesses voids, the structures of the three thinnest prisms (T/L = 1/50, 1/25 and 1/10, which exhibit polarity) is comprised of long, nearly linear runs of triangles bound in close-packed edge-edge states (Fig. 14a – c). By contrast, the thickest prisms (T/L = 1/5, which do not exhibit polarity) contain fewer close-packed prisms, and no linear chains of edge-edge bonds (Fig. 14d).

Network porosity is quantitatively assessed by computing a common measure of number density fluctuations:

$$\chi_L = \frac{\langle N^2 \rangle - \langle N \rangle^2}{\langle N \rangle}\bigg|_L$$

Here is N is the number of particles within an ensemble of square bins of size L. The brackets denote the average over the ensemble. This quantity is equivalent to the compressibility in the long wavelength limit; we here refer to it as $\chi_L$. The quantity $\chi_L$ has previously been used to describe the long-range structure of colloidal gels.[43–45] The network images in Fig. 10 were divided into square regions of 240 x 240 μm², 480 x 480 μm², and 720 x 720 μm² and the compressibility measure determined for each bin size; the results are reported in SI Table 3. For each bin size, the compressibility measure is greatest for the networks of the thinnest particles (Figs. 14a-c), progressively decreasing for the network of thicker prisms (Fig. 10d); this quantitative result is consistent with the images in Figure 14, which show larger voids for the thin prism networks relative to the thick networks.

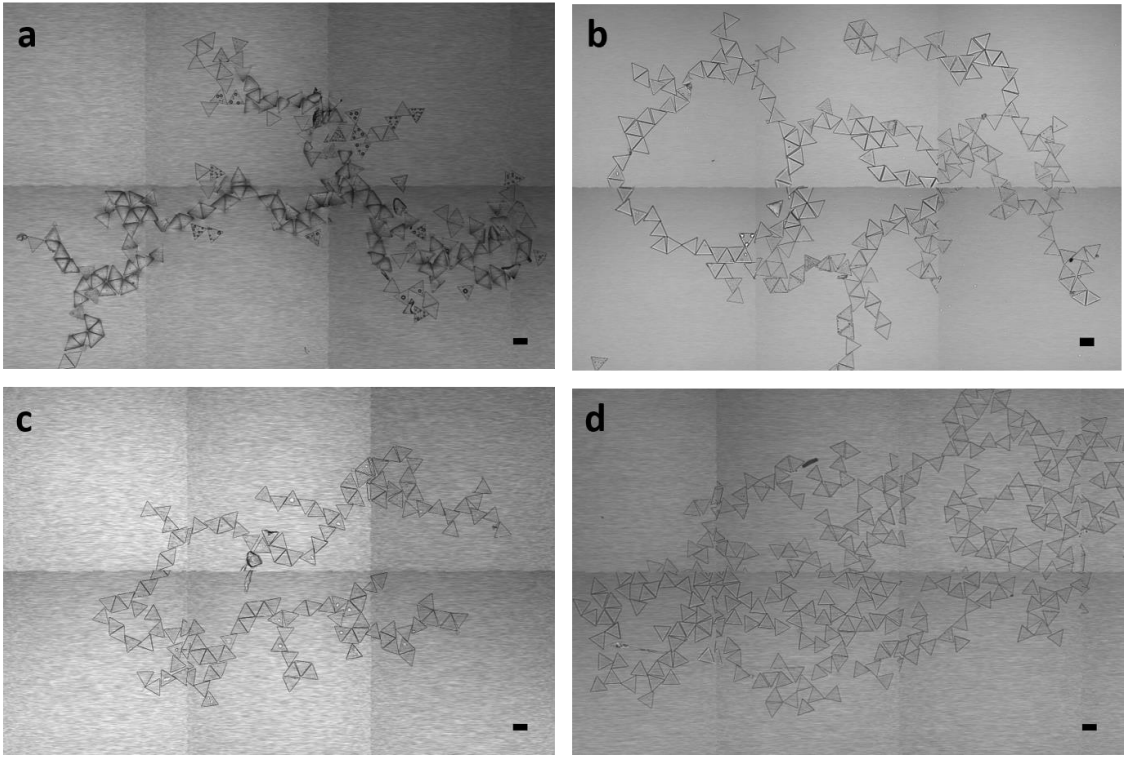

*Figure 14 Self-assembled open networks from capillary-driven binding of thin triangular microprisms. (a) T/L = 1/50, (b) T/L = 1/25, (c) T/L = 1/10, and (d) T/L = 1/5 equilateral triangular microprisms. Scale-bars are 100 μm.*

**Discussion**

In the discussion that follows, we comment on the ramifications of the coupled prism polarity and hexapolar-like interactions of the thinnest prisms. We address how the polarity of pair-binding prisms is predicative of both pair-binding trajectories and of the final pair-bonded state. We then discuss the effect of prism polarity on open network structure and suggest a path to design a prism building block for an ordered kagome lattice.

*Prism polarity is predictive of tip-tip vs. tip-edge binding trajectory*

The results show that for thin prisms (T/L = 1/25 and 1/50) the type of prism-prism bond formed may be predicted with 100% fidelity from the polarity of the two prisms participating in the bonding event. Prisms of the same polarity only access the tip-to-tip trajectory which then leads to the tip-tip and edge-edge final binding states, while prisms of opposite polarity only access the tip-to-midpoint edge trajectory, which then leads to tip-to-midpoint edge and edge-edge offset binding states. Our observations suggest that the tip-tip and tip-edge binding states only survive at steady-state when the collapse of the prisms into their edge-edge or edge-edge offset states is frustrated, due to, either geometrically induced frustration from surrounding prisms or roughness at the prism sides, which prevent rotations. Prism polarity – and its control over prism-prism binding trajectory – is also observed for T/L = 1/10 prisms, although evidence of the effect is not as obvious with optical microscopy (Fig. 4c, columns 2 and 3) and thus was not analyzed in the same way thinner prisms are in Figs. 6 and S3. T/L = 1/5 prisms, on the other hand, lack observable polarity. Fig. 4d shows a variety of bonding states along the edge – instead of localization at the tip and midpoint edge as seen for the thin prisms.

*Hexapole-like capillary interactions from interface-prism contact line bowing*

Due to the flat geometry of the prism sides, instead of an equilibrium contact line with constant contact angle (as in the case of cylinders at an interface), the triangular prisms leads to contact lines pinned at edges of the triangular face, as we discussed above. Bowing of thinner prisms leads to a contact line that is conformal to that of the bowed triangle surface. Thus a hexapole-like interface profile around them arises, wherein tips and edges of the triangle exhibit opposite interface height variations. Interestingly, the interface profile differs from that of ideal hexapoles, especially close to the prisms, due to the focusing of excess interface area near the tips. We find this to be the origin of the tip-to-tip attraction for same polarity prisms, which may be a useful mechanism to obtain regular open networks.

In contrast, we find thicker prisms to be much more flat, and thus the pinned contact line does not exhibit a significant hexapolar component. Instead, it likely generates an interaction, described by Fourier decomposition of the variability in interfacial height profile, as generated by non-ideal features of the flat surface, such as its roughness. This leads to quadrupole-quadrupole interactions at far field, consistent with our observation from the binding events.

*Open network structure and the path to capillary-drive self-assembly of ordered open lattices*

The open networks shown in Fig. 14 display a heterogeneous structure, characterized by disordered strands and voids. These structures are reminiscent – and perhaps even more open than – networks self-assembled from colloidal ellipsoids at fluid-fluid interfaces [7], which demonstrate enhanced rigidity as compared to close-packed arrays of isotropic spheres. Recent theoretical studies show unusual mechanical properties of regular open structures such negative Poisson's ratio in the twisted kagome lattice. Mechanical properties of these disordered open networks will be an interesting direction for future research.[12,14,18,31,33]

Open networks self-assembled from the three thinnest prisms, which exhibit polarity and possess a capillary hexapole, (Figs. 14a-c) are more heterogeneous than are networks self-assembled from the thickest prisms, which exhibit neither polarity nor a capillary hexapole (Fig. 14d). This is demonstrated quantitatively through the measurement of particle number density fluctuations (Fig. S5); open networks self-assembled from the thinnest prisms exhibit higher particle number density fluctuations than do open networks self-assembled from the thickest prisms. These results suggest that future work could understand how void structure in such disordered networks could be controlled by design of building block shape, surface properties, and pair interactions in capillary systems.

On the other hand, to realize regular open networks, such as the kagome lattice, where only tip-tip binding is selected, further studies are needed in order to (1) select a single polarity component of the prisms, and (2) stabilize the binding of the prisms at the tip-tip configuration and avoid the collapse into edge-edge binding. The former may be addressed by introducing Janus character to the particles, so that they attach to the interface in just one of the two possible configurations. The latter may be realized by optimizing the shape of the prisms such that the tips are slightly truncated such that the tip-tip configuration is a local minimum.

**Conclusion**

We have reported capillary-driven binding of thin, triangular prisms of T/L between 1/50 and 1/5 into open networks at a flat air-water interface. The interface pins to the concave face of the three thinnest prisms (T/L = 1/50, 1/25, and 1/10). Interface pinning and physical bowing of the thin prisms results in (a) two polarities corresponding to prism bowing up, interface pinned at top edges, prism center-of-mass below interface, and prisms bowing down, interface pinned at bottom edges, prism center-of-mass above interface, and (b) hexapolar-like interface profile around the prisms. The resulting capillary interactions between these triangular prisms lead to tip-tip, edge-edge, tip-edge, and edge-edge-offset pair binding events, depending on the polarity of the pair, and disordered open networks produced by self-assembly. Thick prisms (T/L = 1/5) exhibit neither physical bowing nor splitting of the prisms into two subpopulations above and below the air-water interface. Prisms of all thicknesses self-assemble into open networks with void structure that depends on the geometric properties of the prism. The results can inform the design of thin prism building blocks for assembly of open networks at fluid-fluid interfaces with either order or disordered structure.

**Conflicts of Interest**

There are no conflicts of interest to declare.

**Acknowledgements**

This work was supported by the National Science Foundation under grant numbers NSF CBET 1232937 and NSF DMR 1609051. For eSEM imaging, we acknowledge the University of Michigan College of Engineering for financial support and the Michigan Center for Materials Characterization for use of the instruments and staff assistance.

**Supplementary Information: Capillary-driven binding of thin triangular prisms at fluid interfaces**

Joseph A. Ferrar, Deshpreet S. Bedi, Shangnan Zhou, Peijun Zhu, Xiaoming Mao,* and Michael J. Solomon*

(Corresponding Author: maox@umich.edu and mjsolo@umich.edu)

**Multipolar interactions between particles at a fluid interface**

In this section we describe equilibrium interface shapes and the resulting capillary interaction between circular multipoles. In the Results section we will show that the capillary interaction between two triangular prisms in our experiment is similar to hexapolar interactions.

The pressure difference across an interface between two stationary, immiscible fluids is given by the Young-Laplace equation,

$$\Delta p = p_1 - p_2 = -\gamma \nabla \cdot \mathbf{n},$$

where $\gamma$ is the surface tension and $\mathbf{n}$ is the unit vector pointing from the lower fluid (2) to the upper fluid (1). Note that $-\nabla \cdot \mathbf{n} = 2H$, where $H$ is the mean curvature of the interface surface. Suppose that the height of the interface is given by $h(\mathbf{x})$, where the far-field equilibrium height of the interface is $h = 0$ (the interface is flat and, consequently, the pressure difference across the interface is zero). We can write the Young-Laplace equation in terms of the height field as

$$\nabla \cdot \frac{\nabla h}{\sqrt{1 + |\nabla h|^2}} = \kappa^2 h$$

where

$$\kappa = \ell_c^{-1} \equiv \sqrt{\frac{(\rho_2 - \rho_1)g}{\gamma}}$$

is the inverse capillary length. The capillary length is a characteristic length scale arising from comparing the relative strengths of gravitational acceleration and the surface tension; for length scales much smaller than the capillary length, the effects of gravity can be neglected. The capillary length of an air-water interface is 2.7 mm.

We can simplify the governing equation of the interface height $h$ by making two assumptions that are typically satisfied by micron-sized particles. First, the interface slope is taken to be small: $|\nabla h|^2 \ll 1$. Second, we consider length scales that are much smaller than the capillary length, $l \ll \ell_c$. In this case, the Bond number is vanishingly small, $\text{Bo} = (\kappa l)^2 \ll 1$, and the Young-Laplace equation simplifies to the 2-D Laplace's equation,

$$\nabla^2 h = 0. \tag{1}$$

Let us consider the case of a solid particle adsorbed to the interface such that the contact line between the interface and the particle surface is undulating. This can be due to particle shape (anisotropies, corners, and edges) and surface roughness/irregularities. These undulations can be decomposed into a multipole expansion such that this differential equation can be solved analytically, for particles with circular cross-sections, using polar coordinates $(r, \theta)$. The solution for the interface height profile is

$$h(r,\theta) = H_0 \ln(\kappa r) + \sum_{m=1}^{\infty} H_m \left(\frac{r_0}{r}\right)^m \cos[m(\theta - \theta_{m,0})], \qquad (2)$$

where $H_m$ is the amplitude of the $m$th moment at the surface/circumference of the particle's circular projection, $r_0$. $m \in \mathbb{Z}^+ \cup \{0\}$ is the multipole moment, and $m = 0,1,2,3$ correspond to the monopole/charge, dipole, quadrupole, and hexapole moments, respectively. If the particle adsorbed to the interface is sufficiently light, the monopole moment vanishes; if the particle is allowed to spontaneously rotate about a horizontal axis, then the dipole moment also vanishes. Therefore, the quadrupole moment ($m = 2$) is typically the leading non-zero term in the multipole expansion (Figure ).

For two particles with circular cross-sections, it is convenient to use bipolar coordinates $(\omega, \tau)$ to obtain a solution to Eq.(1). They are defined implicitly via $x = a\dfrac{\sinh\tau}{\cosh\tau - \cos\omega}$, $y = a\dfrac{\sin\omega}{\cosh\tau - \cos\omega}$, where $\tau \in \mathbb{R}$, $\omega \in [0, 2\pi)$ (or, equivalently, $\omega \in [-\pi, \pi)$). Curves of constant $\omega$ and $\tau$ are circles that intersect at right angles in the $xy$-plane.

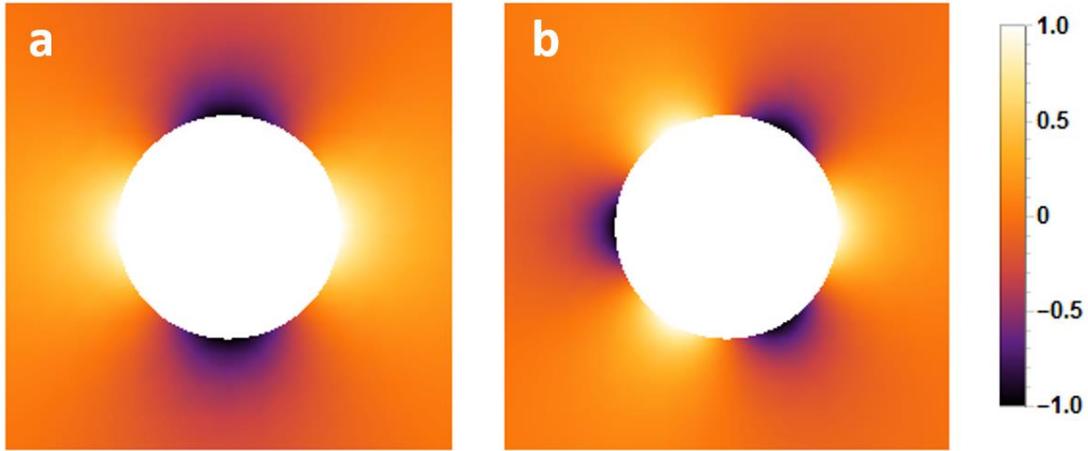

*Figure S1. Theoretical interface height profile for particles with circular cross-sections. (a) A capillary quadrupole ($m = 2$), with four alternating regions of positive and negative interface height (the equilibrium interface height far from any particles is taken to be zero), and (b) A capillary hexapole ($m = 3$) with six alternating regions of positive and negative interface height.*

The parameter $a$ is determined by the particle radii and their separation distance

$$a^2 = \frac{1}{4r^2}[r^2 - (R_1 + R_2)^2][r^2 - (R_1 - R_2)^2].$$

Note that, in the bipolar coordinate system, the circular projections of the contact lines on the $xy$-plane are curves of constant $\tau$, $\tau = -\tau_1$ and $\tau = \tau_2$, where

$$\tau_1 = \cosh^{-1}\left(\frac{r^2 + R_1^2 - R_2^2}{2rR_1}\right), \qquad \tau_2 = \cosh^{-1}\left(\frac{r^2 + R_2^2 - R_1^2}{2rR_2}\right)$$

Rewriting Laplace's equation in terms of bipolar coordinates ultimately yields a deceptively simple partial differential equation of the form

$$\frac{\partial^2 h}{\partial \omega^2} + \frac{\partial^2 h}{\partial \tau^2} = 0.$$

The derivation can be found in Ref. X\[1] the resultant interface solution is

$$\begin{aligned}
h(\omega, \tau) &= H_1 \sum_{n=1}^{\infty} A(n, m_1, \tau_1) \cos(n\omega - m_1 \theta_1) \frac{\sinh[n(\tau_2 - \tau)]}{\sinh[n(\tau_1 - \tau_2)]} \\
&+ H_2 \sum_{n=1}^{\infty} A(n, m_2, \tau_2) \cos(n\omega - m_2 \theta_2) \frac{\sinh[n(\tau_1 + \tau)]}{\sinh[n(\tau_1 + \tau_2)]}
\end{aligned}$$

where

$$A(n, m_i, \tau_i) = m_i \sum_{k=0}^{\min(m_i, n)} \frac{(-1)^{m_i - k}(m_i + n - k - 1)!}{(m_i - 1)!(n - k)!\, k!} \exp[-(m_i + n - 2k)\tau_i].$$

*Interaction potential between two capillary multipoles*

The capillary interaction potential between two particles is a function of their orientations and separation distance. It is given by

$$U_{12} = \gamma(\delta S_{12} - \delta S_1 - \delta S_2), \tag{3}$$

where $\delta S_{12}$ is the excess area created at the interface in the full two-particle system, and $\delta S_i$ ($i = 1,2$) is the excess area in an isolated one-particle system (i.e., the separation distance $r \to \infty$). The excess area is defined as the difference between the actual surface area $\Sigma^*$ and the projected surface area $\Sigma$ (the interface would be planar without the deformation caused by the particle) [2] In the small slope regime, the excess surface area is given by

$$\delta S = \frac{1}{2} \iint_\Sigma dS\, |\nabla h|^2.$$

From these two preceding equations, it is apparent that minimization of the capillary interaction potential coincides with the minimization of excess area beyond that created by two isolated particles. This favors the adoption of particle configurations such that the slope of the resultant interface is reduced. For particles with fixed orientations, the interaction between the two will be attractive if moving the particles closer together will reduce the overall slope of the interface (and, thus, decrease the amount of excess interfacial area) and repulsive if moving the particles further apart will reduce the overall slope.

For a single particle with a circular cross-section and a contact line that is undulating with multipole moment $m$, the excess surface area is [1]

$$\delta S_i = \frac{\pi}{2} m_i H_i^2.$$

For two capillary multipoles, the excess surface area is

$$\delta S_{12} = \pi [H_1^2 S_1 + H_1^2 S_2 - H_1 H_2 G \cos(m_2 \theta_2 - m_1 \theta_1)]$$

where

$$S_i = \sum_{n=1}^{\infty} \frac{n}{2} \coth[n(\tau_1 + \tau_2)] A^2(n, m_i, \tau_i)$$
$$G = \sum_{n=1}^{\infty} \frac{n A(n, m_1, \tau_1) A(n, m_2, \tau_2)}{\sinh[n(\tau_1 + \tau_2)]}$$

and

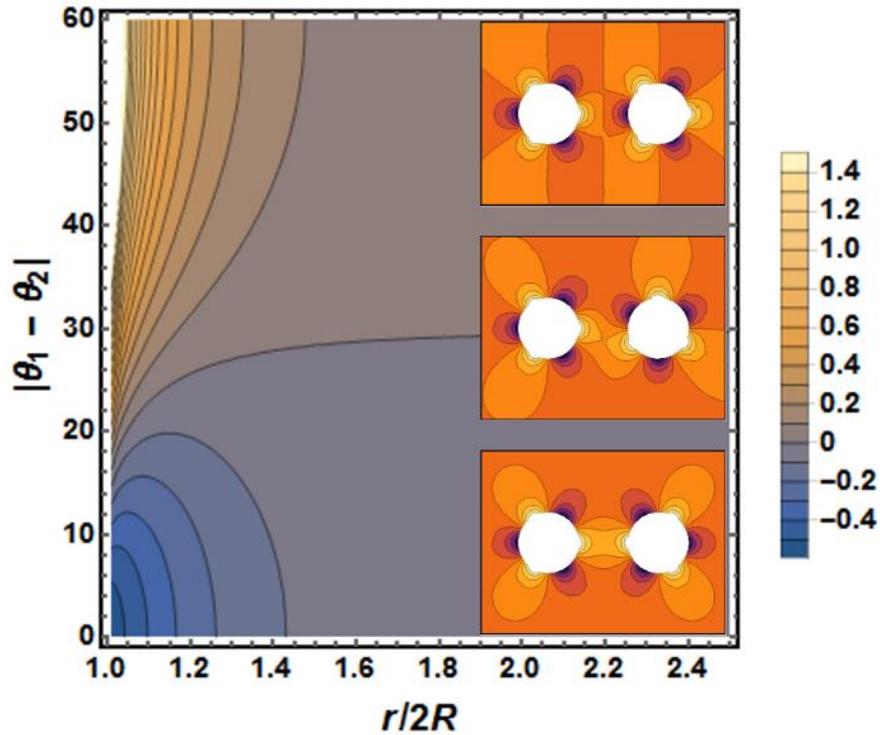

*Figure S2 Theoretical capillary interaction potential between two capillary hexapoles as a function of separation distance, $r$, scaled by the diameter of the particles' circular projection, $2R$, and the particles' relative orientation, $|\theta_1 - \theta_2|$. The three insets show the interface height profile of three configurations corresponding to relative orientations of 0°, 30°, and 60° at a distance of $r/2R = 1.8.$.*

$$A(n, m_i, \tau_i) = m_i \sum_{k=0}^{\min(m_i,n)} \frac{(-1)^{m_i-k}(m_i + n - k - 1)!}{(m_i - 1)!\,(n - k)!\,k!} \exp[-(m_i + n - 2k)\tau_i].$$

Here, it is important to realize that, for two c888apillary multipoles of the same order, such that $m_1 = m_2$, the interaction energy reduces to a two-dimensional function of their separation distance, $r$, and their *relative* orientation, $|\theta_1 - \theta_2|$. An example of hexapole-hexapole interaction energy is shown in Figure .

*Contact-line boundary conditions*

The solution to the Young-Laplace equation is subject to two boundary conditions: one at the three-phase (solid, liquid, and fluid, with the latter oftentimes a gas) contact line and one at the far boundary of the interfce, infinitely far away. The latter is typically taken to be the condition of a flat interface. The boundary condition at the contact line, however, can be more complicated. In the simplest case, in which the surface of the solid phase (e.g., a wall or a particle) is energetically homogeneous, the contact line is determined such that the *equilibrium* contact angle, $\theta_c$, between the solid surface and the surface of the interface is constant and satisfies the Young equation [3,4]

$$\gamma \cos \theta_c = \gamma_{SG} - \gamma_{SL},$$

where $\gamma, \gamma_{SG}, \gamma_{SL}$ are the liquid-gas, solid-gas, and solid-liquid surface tensions, respectively.

In this paper, due to the specific shape of the particles used in the experiment – triangular prisms – we will focus on a specific boundary condition in which the contact line is kinetically trapped, or pinned, at sharp corners and edges of a particle. This pinning results in a non-equilibrium contact angle that can deviate significantly from the equilibrium contact angle discussed above and can also vary along the contact line. As shown by Gibbs in an extension to the Young equation, [5,6] the non-equilibrium contact angle, $\theta_g$, at a pinned edge can be any value in the range

$$\theta_c \leq \theta_g \leq \pi - \delta + \theta_c,$$

where $\delta$ is the wedge angle of the particle. For instance, the wedge angle of the top or bottom edges of a cube is $\pi/2$. Note that the limiting angles of Gibbs' criterion or inequality are simply the equilibrium contact angles for each of the two surfaces that join together to form the edge with a wedge angle of $\delta$; when $\theta_g$ extends beyond the bounds of the inequality, the contact line becomes unpinned and begins to slide along one of the two surfaces, as dictated by which bound was violated. [7] This phenomenon of contact-line pinning has been observed in various experimental systems consisting of solid particles or substrates containing sharp edges. [8–10] For example, in the case of a small cylindrical particle with negligible Bond number oriented vertically, a preferred equilibrium contact angle of $\theta_c \neq \pi/2$ cannot be achieved anywhere along the side of the cylinder; therefore, the contact line will either move up (if the preferred contact angle $\theta_c < \pi/2$) or down (if $\theta_c > \pi/2$) until either the top or bottom face, respectively, of the cylinder coincides with the interface.[11] In this case, the contact line is pinned to the edge of the cylinder with non-equilibrium contact angle $\theta_g = \pi/2$, and the surrounding interface is completely planar.

In this experiment, the equilibrium contact angle of the air-water interface with the triangular prisms has been measured to be about 5 degrees, and the prisms have a wedge angle of 90 degrees (at both the top and the bottom). For these specific values, Gibbs' criterion ostensibly implies that mechanical equilibrium for a particle of negligible weight can only be satisfied when the contact line is pinned to the edges of the top face of a particle (as it is only here that the range of permissible contact angles allows for both upward- and downward-pointing interface/surface tension vectors such that they sum to zero over the closed contact line loop, as is required by the condition of mechanical equilibrium). Incorporating the fact that the triangular prisms are bowed, however, it can be seen that only downward-pointing interface vectors are possible if the contact line is pinned to the edges of the top face of a bowed-down prism – such a prism is only mechanically stable when the contact line is pinned to the edges of the *bottom* face instead. It remains true for a bowed-up prism, nevertheless, that the contact line necessarily is pinned to the edges of the top face. In both cases, then, the contact line needs to pin to the edges of the *concave* face of the bowed triangular prism, which is consistent with experimental observation.

**Triangular Prism Binding States Correlate with Prism Polarity**

In Tables 1 (as derived from Figure 6) and 2 (as derived from SI Figure S3), 237 bonds are analyzed across six 1270 x 1270 μm regions of open networks of T/L = 1/25 and 1/50 prisms (three 1270 x 1270 μm regions per network). 17 bonds are between prisms with indeterminate polarity – prisms whose polarity cannot be resolved by optical microscopy – and are not included in this analysis. Of the 220 remaining bonds, there is perfect agreement in the number of bonds between prisms with the same polarity (133 bonds) and prisms bound tip-tip or edge-edge (133 bonds), and there is also perfect agreement in the number of bonds between prisms with the opposite polarity (87 bonds) and prisms bound tip-edge or edge-edge offset (87 bonds).

Averaging over three locations in each network, and counting over the 237 total events, bonds between prisms with the same polarity account for 48% of all bonds for T/L = 1/50 prisms and 64% of all bonds for T/L = 1/25 prisms; bonds between prisms with the opposite polarity account for 37% of all bonds for T/L = 1/50 prisms and 36% of all bonds for T/L = 1/25 prisms, and bonds between prisms with indeterminate polarity account for 15% of all bonds for T/L = 1/50 prisms. Bonds of indeterminate polarity are not observed for T/L = 1/25 prisms.

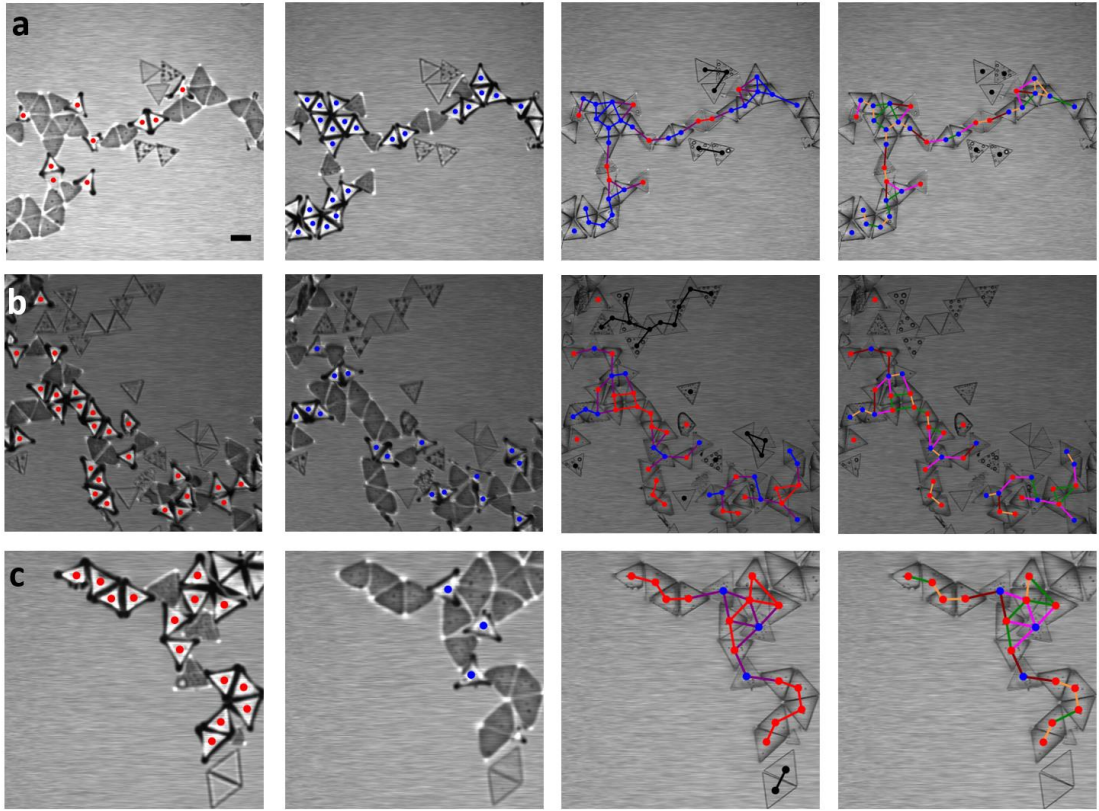

*Figure S3* *Identification of triangular prism binding states for (T/L = 1/50). Each row of images (a) – (c) represents a different location within a network structure. The relative position of microscope's focal plane to the air-water interface is varied by column as follows: Column (1): Microscope focal plane is ~200 μm below the interface. Clearly visible prisms are identified with red markers. Column (2): Microscope focal plane is ~200 μm above the interface. Clearly visible prisms are identified with blue markers. Column (3): Microscope focal plane is at the interface. Bonds between prisms with the same polarity are identified with blue and red connecting lines, bonds between prisms with the opposite polarity are identified with purple connecting lines, bonds between prisms with indeterminate polarity are identified with black connecting lines. Column (4): Microscope focal plane is at the interface. Prism-prism bonds are identified by their polarity-independent orientation: side-side (orange connecting lines), tip-tip (green connecting lines), side-side offset (brown connecting lines), tip-side (pink connecting lines). Bonds in Columns (3) and (4) are tabulated in Table (2). Scale-bar is 100 μm.*

| Fig. No. | Relative polarity of bound prisms | No. of Bonds | % of each type of bond | tip-tip | edge-edge | No. of tip-tip + edge-edge | Correlation between relative polarity of bound prisms and bond orientation | tip-edge | edge-edge offset | No. of tip-edge + edge-edge offset | Correlation between relative polarity of bound prisms and bond orientation |
|---|---|---|---|---|---|---|---|---|---|---|---|
| 5a | Same | 10 | 50% | 8 | 2 | 10 | 100% | 0 | 0 | 0 | 0% |
|  | Opposite | 10 | 50% | 0 | 0 | 0 | 0% | 5 | 5 | 10 | 100% |
|  | Indeterminate | 0 | 0% | - | - | - | - | - | - | - | - |
|  | Total | 20 |  |  |  |  |  |  |  |  |  |
| 5b | Same | 32 | 66% | 19 | 13 | 32 | 100% | 0 | 0 | 0 | 0% |
|  | Opposite | 16 | 34% | 0 | 0 | 0 | 0% | 10 | 6 | 16 | 100% |
|  | Indeterminate | 0 | 0% | - | - | - | - | - | - | - | - |
|  | Total | 48 |  |  |  |  |  |  |  |  |  |
| 5c | Same | 36 | 67% | 24 | 12 | 36 | 100% | 0 | 0 | 0 | 0% |
|  | Opposite | 18 | 33% | 0 | 0 | 0 | 0% | 15 | 3 | 18 | 100% |
|  | Indeterminate | 0 | 0% | - | - | - | - | - | - | - | - |
|  | Total | 54 |  |  |  |  |  |  |  |  |  |
| Fig. 5 Total | Same | 78 | 64% | 51 | 27 | 78 | 100% | 0 | 0 | 0 | 0% |
|  | Opposite | 44 | 36% | 0 | 0 | 0 | 0% | 30 | 14 | 44 | 100% |
|  | Indeterminate | 0 | 0% |  |  |  |  |  |  |  |  |
|  | Total | 122 |  |  |  |  |  |  |  |  |  |

*Table S1. Comparison of prism-prism bond type based on polarity of bound prisms and polarity-independent prism orientation for T/L = 1/25. All data is tabulated from analysis described in Fig. 5. Bonds are sorted into rows by the relative polarity of the bound prisms (same, opposite, or indeterminate polarity and into columns by the polarity-independent orientation of the bound prisms. The correlation between the relative polarity of the bound prisms and the polarity-independent bond orientation is calculated for network location analyzed. All bond types and correlations are also totaled over all 3 network locations.*

| Fig. No. | Relative polarity of bound prisms | No. of Bonds | % of each type of bond | tip-tip | edge-edge | No. of tip-tip + edge-edge | Correlation between relative polarity of bound prisms and bond orientation | tip-edge | edge-edge offset | No. of tip-edge + edge-edge offset | Correlation between relative polarity of bound prisms and bond orientation |
|---|---|---|---|---|---|---|---|---|---|---|---|
| 6a | Same | 23 | 58% | 10 | 13 | 23 | 100% | 0 | 0 | 0 | 0% |
|  | Opposite | 14 | 35% | 0 | 0 | 0 | 0% | 8 | 6 | 14 | 100% |
|  | Indeterminate | 3 | 8% | - | - | - | - | - | - | - | - |
|  | Total | 40 |  |  |  |  |  |  |  |  |  |
| 6b | Same | 20 | 41% | 9 | 11 | 20 | 100% | 0 | 0 | 0 | 0% |
|  | Opposite | 20 | 41% | 0 | 0 | 0 | 0% | 14 | 6 | 20 | 100% |
|  | Indeterminate | 13 | 27% | - | - | - | - | - | - | - | - |
|  | Total | 53 |  |  |  |  |  |  |  |  |  |
| 6c | Same | 12 | 48% | 6 | 6 | 12 | 100% | 0 | 0 | 0 | 0% |
|  | Opposite | 9 | 48% | 0 | 0 | 0 | 0% | 5 | 4 | 9 | 100% |
|  | Indeterminate | 1 | 5% | - | - | - | - | - | - | - | - |
|  | Total | 22 |  |  |  |  |  |  |  |  |  |
| Fig. 6 total | Same | 55 | 48% | 25 | 30 | 55 | 100% | 0 | 0 | 0 | 0% |
|  | Opposite | 43 | 37% | 0 | 0 | 0 | 0% | 27 | 16 | 43 | 100% |
|  | Indeterminate | 17 | 15% |  |  |  |  |  |  |  |  |
|  | Total | 115 |  |  |  |  |  |  |  |  |  |

*Table S2. Comparison of prism-prism bond type based on polarity of bound prisms and polarity-independent prism orientation for T/L = 1/50. All data is tabulated from analysis described in Fig. 6. Bonds are sorted into rows by the relative polarity of the bound prisms (same, opposite, or indeterminate polarity and into columns by the polarity-independent orientation of the bound prisms. The correlation between the relative polarity of the bound prisms and the polarity-independent bond orientation is calculated for network location analyzed. All bond types and correlations are also totaled over all 3 network locations.*

**Pair Binding Trajectories**

Movies S1-S7 show pairwise binding trajectories analyzed in main text Fig. 10 and 11, and in supplementary Fig. S4.

Movie S1: T/L = 1/50 tip-tip, frame rate = 30 fps (Fig. S4a)

Movie S2: T/L = 1/50 tip-edge, frame rate = 30 fps (Fig. S4b)

Movie S3: T/L = 1/25 tip-tip trial 1, frame rate = 15 fps

Movie S4: T/L = 1/25 tip-tip trial 2, frame rate = 15 fps (Fig. S4c)

Movie S5: T/L = 1/25 tip-tip trial 3, frame rate = 15 fps

Movie S6: T/L = 1/25 tip-edge trial 1, frame rate = 15 fps

Movie S7: T/L = 1/25 tip-edge trial 2, frame rate = 15 fps (Fig. S4d)

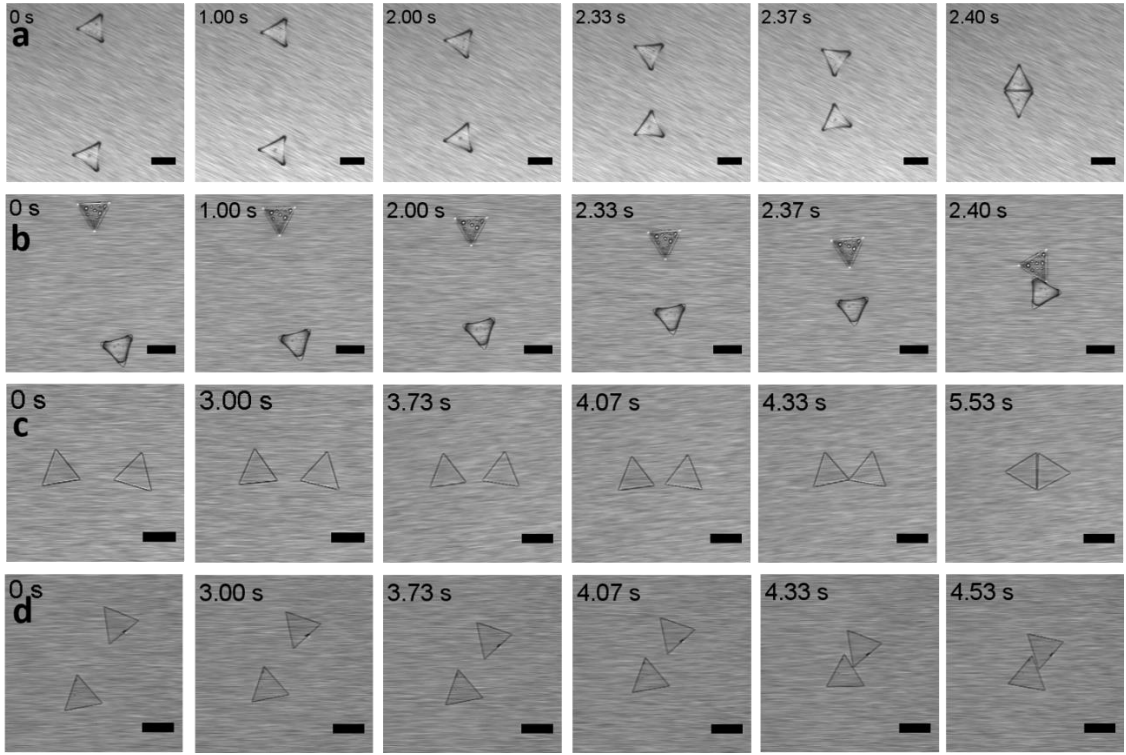

*Figure S4 Optical microscopy images of the 2 types of binding trajectories observed for polar prisms (T/L ≤ 1/10), shown for T/L = 1/50 (rows (a) and (b)) and T/L = 1/25 (rows (c) and (d)). For prisms of T/L = 1/50 (rows (a) and (b)), contact occurs between the 5$^{th}$ and 6$^{th}$ images of each row. For prisms of T/L = 1/25 (rows (c) and (d)), contact occurs in the 5$^{th}$ image of each row. Rows (a) and (c), tip-to-tip binding trajectory: the prisms approach and first contact occurs at the tips. The prisms then rotate into a collapsed, fully flush edge-to-edge orientation. Rows (b) and (d), tip-to-midpoint edge binding trajectory: the prisms approach and contact one another in an orientation such that the tip of one prism binds at the midpoint of the other prism's edge. The prisms then rotate*

**Heterogeneity of Self-Assembled Networks**

For regions of 240 x 240 $\mu m^2$, networks of the three thinnest particles have a compressibility measure of 2.6 ± 0.5, while the network of the thickest particles has a number density fluctuation of 1.8 ± 0.1. For regions of 480 x 480 $\mu m^2$, networks of the three thinnest particles have a number density fluctuation of 5.8 ± 0.6, while the network of the thickest particles has a number density fluctuation of 4.4 ± 0.2. For regions 720 x 720 $\mu m^2$, networks of the three thinnest particles have a number density fluctuation of 10.4 ± 0.9, while the network of the thickest particles has a number density fluctuation of 7.7 ± 0.2.

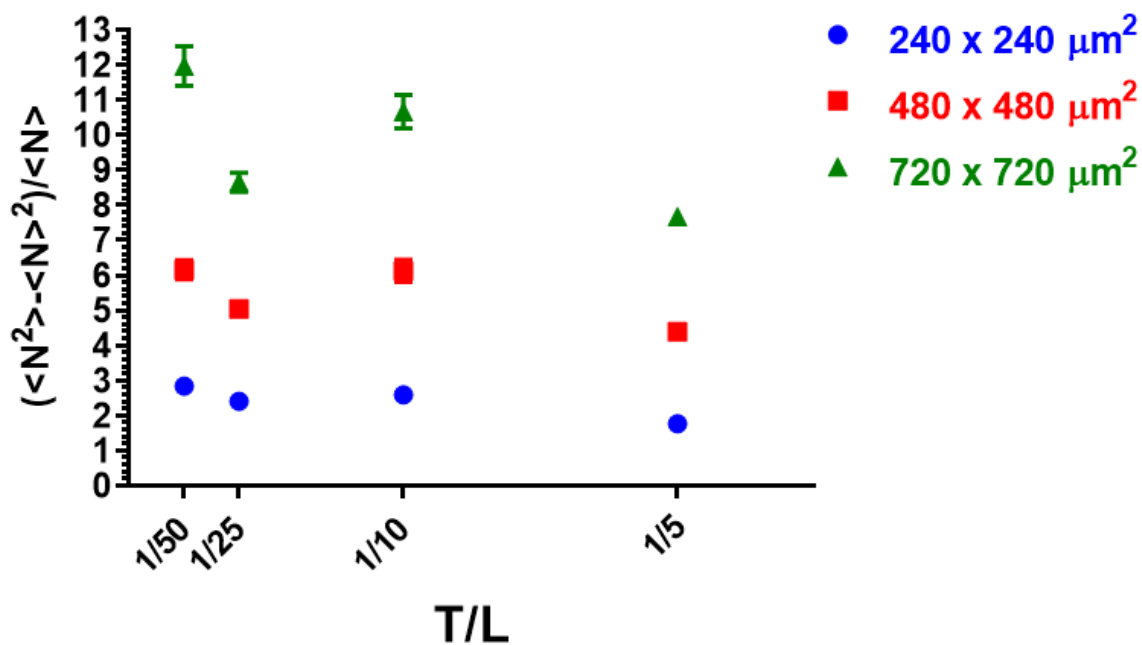

*Figure S5. Mean-squared particle number density fluctuation for the networks shown in Figure 14.*

**Additional simulated pair-binding trajectories**

Three representative initial conditions, corresponding to configurations close to (but purposefully not exactly) tip-to-tip, tip-to-side, and side-to-side were selected, and the resultant simulated trajectories are shown in Figure 15 for two different viscous-damping coefficient ratios, $\eta_{\tilde{\theta}}/\eta_r = 1.46, 0.146$. In all three sets of trajectories, it is clear that mirror symmetric configurations are preferred -- for cases where the initial configuration is already mirror symmetric, the subsequent configurations remain mirror symmetric; otherwise, the particles will first rotate to a mirror symmetric configuration. For smaller ratios and fixed $\eta_r$, $\eta_{\tilde{\theta}}$ becomes correspondingly smaller, meaning that it is easier for the particles to rotate. This accounts for the fact that, in all cases, the $\theta_1 = \theta_2 = 0°$ tip-tip mirror-symmetric configuration is more easily achieved for the smaller ratio value.

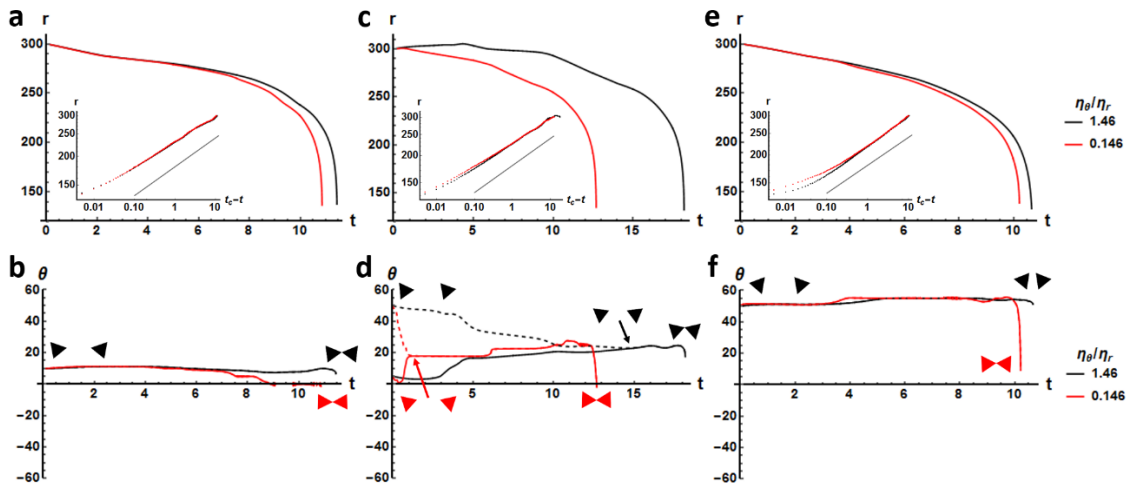

*Figure S6 Configuration trajectories for three representative initial conditions (close to (a),(b) tip-to-tip, (c),(d) tip-to-side, and (e),(f) side-to-side) and two different viscous-damping coefficient ratios. The top row shows the separation distance as a function of simulation time, with insets plotting separation distance values as a function of time-to-contact on a log scale. The gray reference line corresponds to the theoretical case of two ideal hexapoles approaching each other in a mirror-symmetric configuration. The bottom row shows the orientation angles of the triangular prisms as a function of simulation time.*